\documentclass[lettersize,journal]{IEEEtran}
\usepackage[margin=1in]{geometry}
\usepackage{amsmath,amssymb,amsthm, amsfonts}
\usepackage{booktabs}
\usepackage{graphicx} 
\usepackage{subcaption}
\usepackage{xcolor}
\usepackage{tikz}
\usepackage{float}
\usepackage[hidelinks]{hyperref}
\usepackage{algorithmic, algorithm}
\usepackage{array}
\usepackage{textcomp}
\usepackage{stfloats}
\usepackage{url}
\usepackage{verbatim}
\usepackage{multirow}
\usepackage{bbm}
\newtheorem{proposition}{Proposition}

\newtheorem{definition}{Definition}
\newtheorem{remark}{Remark}

\usepackage[style=ieee, citestyle=numeric-comp]{biblatex}
\renewcommand{\algorithmicrequire}{\textbf{Input:}}
\newcommand{\q}{\mathbf{q}}

\newcommand{\A}{\mathbf{A}}
\newcommand{\B}{\mathbf{B}}
\newcommand{\C}{\mathbb{C}}
\newcommand{\I}{\mathbf{I}}
\newcommand{\X}{\mathbf{X}}
\newcommand{\Y}{\mathbf{Y}}
\newcommand{\W}{\mathbf{W}}
\newcommand{\V}{\mathbf{V}}
\newcommand{\U}{\mathbf{U}}

\newcommand{\Z}{\mathbf{Z}}
\newcommand{\Range}{\operatorname{Range}}

\newcommand{\diag}{\operatorname{diag}}

\newcommand{\cD}{\mathcal{D}}
\newcommand{\bDelta}{\boldsymbol{\Delta}}

\begin{document}

\title{Model Order Detection with Parameter Estimation Guarantees for the Matrix Pencil Method}

\author{Yehonatan-Itay Segman\textsuperscript{1,$\dagger$},
        David Sadka\textsuperscript{2,$\dagger$},
        and Ronen Talmon\textsuperscript{1}%
\thanks{\textsuperscript{$\dagger$}Contributed equally.}%
\thanks{\textsuperscript{1}Viterbi Faculty of Electrical and Computer Engineering, Technion -- Israel Institute of Technology, Haifa 32000, Israel; yehonatans@campus.technion.ac.il, ronen@ee.technion.ac.il.}%
\thanks{\textsuperscript{2}Department of Electrical and Computer Engineering, Yale University, New Haven, CT, USA; david.sadka@yale.edu.}%
\thanks{This work has been submitted to the IEEE for possible publication. Copyright may be transferred without notice, after which this version may no longer be accessible.}

}

\markboth{}%
{Shell \MakeLowercase{\textit{et al.}}: A Sample Article Using IEEEtran.cls for IEEE Journals}


\maketitle

\begin{abstract}
Detecting the model order and estimating the parameters 
of sums of damped complex exponentials in noise is challenging when poles are closely spaced, the sample size is small, or the signal-to-noise ratio (SNR) is low. We propose REECAP, a matrix pencil (MP) algorithm for joint model order detection and pole estimation. The algorithm computes candidate poles using an overestimated working rank and classifies them using a score derived from their associated eigenvectors. The model order is then estimated by counting the retained candidates.
The score measures the eigenvector energy concentration in coordinates corresponding to the dominant right singular subspace of the data Hankel matrix. Under a suitable rank condition, we derive a deterministic sufficient condition guaranteeing that a candidate whose score exceeds a certain threshold lies within a prescribed tolerance of at least one signal pole. 
This establishes a theoretical connection between eigenvector energy concentration and pole-estimation accuracy. 
Extensive simulations demonstrate improved model order detection and frequency estimation performance relative to conventional MP order detection methods, recent structure-aware MP methods, and information-theoretic criteria (ITC), across a broad range of scenarios, including clustered frequencies, low SNR in the threshold region, and small sample sizes. Runtime comparisons further demonstrate substantial computational savings, and moderate sensitivity to noise model mismatch, compared to the evaluated structure-aware and information-theoretic methods.
\end{abstract}

\begin{IEEEkeywords}
matrix pencil, super-resolution, model order detection, estimation-accuracy guarantees.
\end{IEEEkeywords}
\begin{refsection}
\section{Introduction}
\label{sec:introduction}
\IEEEPARstart{P}{arametric} spectral estimation is a fundamental problem in signal processing, with applications in radar, telecommunications, audio, and biomedical signal analysis~\cite{percival1993spectral, stoica2005spectral}. 
A common signal model is a finite sum of damped complex exponentials,
\begin{equation}
    x(n) = \sum_{k=1}^{K} b_k z_k^n,
    \qquad
    z_k = e^{-\alpha_k + \jmath \theta_k},
    \label{eq:clean_signal_model}
\end{equation}
where $b_k\in\C$ are nonzero complex amplitudes, $-\pi\leq \theta_k<\pi$ are angular frequencies, and $\alpha_k\geq 0$ are damping factors. The complex numbers $\{z_k\}_{k=1}^{K}$ are termed signal poles. Assuming the signal poles are distinct, their number $K$ defines the model order.
Given \(N\) noisy observations,
\begin{equation}
    y(n) = x(n) + w(n), \qquad n=0,\ldots,N-1,
    \label{eq:noisy_signal_model}
\end{equation}
where $w(n)$ is additive measurement noise, the problem consists of determining the model order, or equivalently the number of signal poles, and estimating their associated parameters defined in~\eqref{eq:clean_signal_model}.

With a finite number of observations, classical Fourier-based
methods have limited resolution, making signal poles with closely spaced frequencies difficult to resolve~\cite{porat1996course}.
Super-resolution methods can resolve signal poles whose frequency separation is smaller than the classical Fourier resolution limit under suitable noise conditions~\cite{krim1996two}.
Parametric super resolution approaches include subspace-based methods such as multiple signal classification (MUSIC)~ \cite{schmidt1986multiple}, estimation of signal parameters via rotational invariance techniques (ESPRIT)~\cite{roy1989esprit}, the MP method~\cite{hua1988matrix}, and the method of direction estimation (MODE)~\cite{stoica1990maximum}.
Subspace-based methods typically require the model order as an input, although it is often unknown. Underestimating it yields biased estimates, while overestimating it can introduce spurious components arising from noise~\cite{papy2007shift}. 
In the ideal noise-free setting, the model order is determined directly from the rank of a data matrix constructed from the observations. With measurement noise, this rank structure is obscured, and the model order must be estimated from the data.

We focus on the MP method, developed by Hua and Sarkar~\cite{hua1988matrix, hua1990matrix}, which has received considerable attention for its superiority over classical Prony-based methods~\cite{hua1988matrix, sarkar1995using}, and its close algebraic connection to ESPRIT~\cite{hua1991svd, potts2013parameter}.
Conventional MP first estimates the model order from a Hankel matrix constructed from the observations. The estimated order then determines the dimension of a generalized eigenvalue problem whose eigenvalues estimate the signal poles~\cite{sarkar1995using}.
The model order is typically estimated by inspecting the singular values of the Hankel matrix, and a reliable detection depends on a clear separation between signal and noise singular values. This separation deteriorates in the low-SNR threshold regime and for closely spaced signal poles~\cite{thomas1995probability,liao2016music}.
An alternative approach for estimating the model order relies on ITC~\cite{stoica2005spectral}, such as the Akaike Information Criterion (AIC)~\cite{akaike1974new} and Minimum Description Length (MDL)~\cite{rissanen1978modeling}.
However, ITC require an explicit statistical model for the observations and incur additional computational cost when evaluated by fitting each candidate model order.

Although several MP variants have been proposed~\cite{hua1988matrix,sarkar1995using,hu1993utilization,lu1998improved, razavilar1998structured}, they primarily focus on improving parameter estimation in noise. These methods retain the conventional sequence of first determining the model order and then estimating the signal parameters from the computed eigenvalues. 
As shown in~\cite{segman2026structure}, this sequence does not exploit the additional information contained in the associated eigenvectors during the detection stage.
The recent structure-aware MP (SAMP) algorithms~\cite{segman2026structure,segman2025samp++} address this limitation by formulating model order detection as an \emph{a posteriori} binary classification of the computed eigenvalues. They choose a working rank \(\widehat K\) larger than the unknown model order \(K\), solve the corresponding \(\widehat K\)-dimensional MP eigenvalue problem, and use structural information contained in the associated right eigenvectors to distinguish signal-related from spurious eigenvalues. The model order is then estimated by counting the signal-related eigenvalues. However, this classification step alone does not ensure the accuracy of the retained signal-related eigenvalues.

In this paper, we propose REECAP, an MP algorithm that jointly determines the model order and identifies computed eigenvalues that provide reliable estimates of the signal poles. 
Building on the a posteriori classification framework of SAMP, we choose a working rank \(\widehat K\) that exceeds the unknown model order \(K\), yielding \(\widehat K\) computed eigenvalues and their associated right eigenvectors. 
We propose a new score that allows us to classify these eigenvalues as true or spurious, where a computed eigenvalue is defined as true if it lies within a prescribed tolerance of at least one signal pole, and as spurious otherwise. 
The score measures the concentration of eigenvector energy in the dominant right singular subspace of the data. For a normalized eigenvector, it reduces to the sum of its dominant squared entries.

Our main theoretical result connects this score to pole estimation accuracy: we show that under a suitable rank condition, there exists a threshold such that any eigenvalue whose score exceeds it is guaranteed to lie within a prescribed tolerance of at least one signal pole.
This establishes a deterministic sufficient condition for accurate pole estimation and provides a theoretical foundation for classifying computed eigenvalues using the proposed score. 
Scoring and classification steps preserve the overall \(\mathcal{O}(N^3)\) complexity of the conventional MP method. Connections to other a posteriori classification methods are discussed in Section~\ref{subsec:prior_works}.

We evaluate REECAP across a broad range of operating conditions, including SNR, sample size, model order, damping, and frequency separation. The results show that it outperforms both conventional model order detection techniques used with the MP method and the more recent SAMP approach, particularly in challenging conditions with closely-spaced frequencies and low SNR values in the threshold region, where model order estimation becomes challenging.
In addition, compared with conventional ITC-based methods, REECAP demonstrates competitive robustness to noise model mismatch while achieving substantially lower computational complexity.

\paragraph*{Organization} 
Sec.~\ref{sec: background} reviews the classical MP method.  In Sec.~\ref{sec: component definition}, we formally define true and spurious computed eigenvalues. Sec.~\ref{sec: proposed score} presents the proposed score and classification algorithm, together with the associated threshold and its theoretical guarantee. Sec.~\ref{subsec:prior_works} discusses the relationship of the proposed method to other methods.
Finally, Sec.~\ref{sec:results} presents simulation results comparing the proposed method with existing baselines.

\paragraph*{Notation}
Throughout, $\|\cdot\|$ denotes the Euclidean norm for vectors and the spectral norm for matrices, with the subscript $2$ omitted for brevity. For a matrix $\mathbf A$, $\sigma_k(\mathbf A)$ denotes its $k$th singular value, in decreasing order. The superscripts $(\cdot)^{\mathrm H}$ and $(\cdot)^\dagger$ denote the conjugate transpose and Moore-Penrose pseudoinverse, respectively. The matrix $\mathbf I$ denotes the identity of appropriate dimension.



\section{Background}
\label{sec: background}
\subsection{Matrix Pencil Method}
\label{subsec:background_MP}
The MP method estimates signal parameters by solving a generalized eigenvalue problem based on the available data samples. It begins by constructing an $(N-L) \times (L+1)$ Hankel matrix $\mathbf{Y}$ from the noisy measurements $y(n)$:

\begin{equation}\label{eq:full noisy Hankel matrix}
\mathbf{Y} =
    \begin{bmatrix}
    y(0)   & y(1)       & \cdots & y(L)   \\
    y(1)   & y(2)       & \cdots & y(L+1) \\
    \vdots & \vdots  & \ddots & \vdots \\
    y(N-L-1) & y(N-L)   & \cdots & y(N-1)
    \end{bmatrix}.
\end{equation} 
This matrix can also be recast as
\begin{equation}\label{eq:Y_decompose}
    \Y = \X + \W,
\end{equation}
where $\X$ is the noiseless Hankel counterpart built from the desired signal $x(n)$ in~\eqref{eq:clean_signal_model}, and $\W$ is the noise only Hankel matrix built from the additive noise vector $w(n)$ in~\eqref{eq:noisy_signal_model}.
The parameter $L$, termed the \emph{pencil parameter}, is chosen to satisfy $K \leq L \leq N-K $ for the noiseless MP theorem to hold~\cite{hua1990matrix}. Using perturbation analysis, it was demonstrated that choosing $\frac{N}{3} \leq L \leq \frac{N}{2}$ minimizes the estimator variance~\cite{rao1985perturbation, hua1988perturbation, hua1988matrix, hua1990matrix, djermoune2009perturbation}. For the rest of this paper, we assume that $L$ lies within this optimal range.

Next, two sub-matrices \(\mathbf{Y}_0\) and \(\mathbf{Y}_1\) are formed from \(\mathbf{Y}\) by removing its last and first columns, respectively. Applying the same
column deletions to $\X$ and $\W$ in~\eqref{eq:Y_decompose} gives 
\[
\Y_i = \X_i + \W_i, \quad i=0,1.
\]
The noiseless counterparts, $\X_0$ and $\X_1$, admit the Vandermonde factorizations~\cite{hua1990matrix}
\begin{equation}\label{eq:clean_vandermonde_factorization}
    \X_0=\Z_L\B\Z_R,
    \qquad
    \X_1=\Z_L\B\Z\Z_R,
\end{equation}
where $\B=\operatorname{diag}(b_1,\ldots,b_K)$, $\Z=\operatorname{diag}(z_1,\ldots,z_K)$, and the Vandermonde matrices $\Z_L\in\mathbb C^{(N-L)\times K}$ and $\Z_R\in\mathbb C^{K\times L}$ are given by
\[
    [\Z_L]_{n,k}    = z_k^{n-1}, \quad
    [\Z_R]_{k,\ell} = z_k^{\ell-1}.
\]
Assuming the signal poles are distinct and $K\leq L\leq N-K$, both Vandermonde matrices have rank $K$. Consequently, the noiseless pencil
\begin{equation}\label{eq:clean_pencil_factorization}
    \X_1-\lambda\X_0
    =\Z_L\B(\Z-\lambda\I)\Z_R
\end{equation}
has rank $K$ except when $\lambda = z_k$, where its rank decreases by one. This is the basis of the MP method~\cite{hua1990matrix}.

Consequently, the MP method estimates the signal poles from the eigenvalues of the noisy pencil $\Y_1-\lambda\Y_0$. 
In the presence of noise, $\Y_0$ and $\Y_1$ are generally full rank (which equals $L$), and a denoising step is performed using the singular value decomposition (SVD) of \(\mathbf{Y}_0\),
\begin{equation} \label{eq:SVD_of_Y_0}
    \Y_0 =  \U \mathbf{\Sigma} \V^{\mathrm{H}},    
\end{equation}
where \(\mathbf{U} \in \mathbb{C}^{(N-L) \times L}\) and \(\mathbf{V} \in \mathbb{C}^{L \times L}\) have orthonormal columns, \(\mathbf{\Sigma} = \text{diag}(\sigma_1, \ldots, \sigma_L)\) with $\sigma_1 \geq \sigma_2 \geq \ldots \geq \sigma_L$, and $(\cdot)^{\mathrm{H}}$ denotes the complex conjugate transpose.
Ideally, a rank-$K$ truncation is applied to~\eqref{eq:SVD_of_Y_0}. However, the true model order $K$ is unknown, and a working-rank 
\begin{equation}\label{eq:working_rank_def}
    1 \leq \widehat K \leq L,
\end{equation}
is typically used to compute the rank-$\widehat K$ truncated SVD 
\begin{equation} \label{eq:truncated_SVD_of_Y_0}
    \mathbf{Y}_0 \approx \mathbf{U}_{\widehat K} \mathbf{\Sigma}_{\widehat K} \mathbf{V}_{\widehat K}^{\mathrm{H}},    
\end{equation}
where $\boldsymbol{\Sigma}_{\widehat K}$ consists of the $\widehat K$ largest singular values in $\boldsymbol{\Sigma}$, and $\U_{\widehat K}$ and $\V_{\widehat K}$ consist of the corresponding $\widehat K$ left and right singular vectors, respectively. This step determines the model order and is further discussed in Section~\ref{subsec:MP_model_order_detection}.

Rather than solving the full-dimensional problem directly, the standard MP formulation estimates the signal poles from the rank-$\widehat K$ projected problem
\begin{equation}\label{eq:projected_problem}
    \mathbf U_{\widehat K}^{\mathrm H}
    (\mathbf Y_1-\lambda\mathbf Y_0)
    \mathbf V_{\widehat K} = \boldsymbol{\Sigma}_{\widehat K}(\A-\lambda\I),
\end{equation}
where 
\begin{equation} \label{eq:projected_matrix_def}
    \A := \mathbf{\Sigma}^{-1}_{\widehat K} \mathbf{U}_{\widehat K}^{\mathrm{H}} \mathbf{Y}_1 \mathbf{V}_{\widehat K}\in\mathbb{C}^{\widehat K \times \widehat K},
\end{equation}
assuming that $\sigma_{\widehat K}(\Y_0) >0$.
In the noiseless case, where $\widehat K = K$, it can be shown~\cite{sarkar1995using} that the eigenvalues of $\A$ coincide with the signal poles $\{z_k\}_{k=1}^K$.
Accordingly, an eigenvalue decomposition of the matrix $\A$ is performed:
\begin{equation}\label{eq:reduced_matrix_eigenvalue_eq}
    \A = \mathbf{Q}{\mathbf{\Lambda}} \mathbf{Q}^{-1}, \quad
    \A\q_i =\lambda_i\q_i, \quad \|\q_i\|=1,
\end{equation}
where ${\mathbf{\Lambda}} = \text{diag}(\lambda_1,\ldots,\lambda_{\widehat K})$ and $\mathbf{Q}\in\mathbb{C}^{\widehat K\times  \widehat K}$ contains the normalized eigenvectors $\mathbf q_i$ as columns.
The estimates of the damping factors and angular frequencies are given by:
\begin{equation}\label{eq:params calc}
    \hat{\alpha}_k = -\text{log}|{\lambda}_k| \,\,, \,\, \hat{\theta}_k = \text{arg}( {\lambda}_k)
    \,\,, k=1,2,\ldots,\widehat{K},
\end{equation}
and the amplitudes $\{b_k\}_{k=1}^{\widehat K}$ are estimated by solving the linear least-squares problem \cite{sarkar1995using}:
\begin{equation}
    \widehat{\mathbf{b}} = \underset{\bar{{\mathbf{b}}}\in\mathbb{C}^{\widehat K}} {\text{arg}\min}\sum\limits_{n=0}^{N-1} |y(n) -  \sum\limits_{k=1}^{\widehat K}{\lambda}_k^n \bar{{b}}_k|^2,
\end{equation} 
where $\widehat{\mathbf{b}} = [\hat{b}_1, \hat{b}_2, \ldots,  \hat{b}_{\widehat K} ]^{\mathrm{T}} \in \mathbb{C}^{\widehat{K}}$.

\subsection{Conventional Model Order Detection Methods}\label{subsec:MP_model_order_detection}
The standard approach for determining the model order in the MP method involves truncating the singular values from the SVD of $\mathbf{Y}_0$ in \eqref{eq:SVD_of_Y_0}, and is a decoupled detection-estimation approach. A commonly used technique is the Significant Decimal Digit (SDD) \cite{hua1990matrix, hua1988matrix, sarkar1995using, laroche1993use, bhuiyan2012advantages}, where the model order is estimated by:

\begin{equation}
    \widehat{K}_{\text{SDD}} = \underset{i=1,\ldots,L}{\text{arg}\min} \; \{\sigma_i : \sigma_i / \sigma_{\text{max}} \geq 10^{-s} \},
\end{equation} 
where $s$ denotes the number of significant decimal digits in the data, $\sigma_{\text{max}}$ is the maximal singular value, and the singular values are assumed to be in descending order.
Another closely related technique is the ``gap'' technique \cite{del1996comparison, yin2011model}, where the model order is estimated by:
\begin{equation}
    \widehat{K}_{\text{Gap}}= \underset{i=1,\ldots,L-1}{\text{arg}\max} \; \left\{\sigma_i /\sigma_{i+1} \right\}.
\end{equation} 

Information-theoretic criteria (ITC) \cite{stoica2004model, stoica2005spectral, nadler2011model}, though less mentioned in the MP literature, can also be used:
\begin{equation}
    \widehat{K}_{\text{ITC}} =   \underset{k=1,\ldots,K_{\text{max}}}{\text{arg}\min} \; \{ -\text{log}\left( p_k (\mathbf{y}|\hat{\mathbf{v}}_k )\right) + k C(k,N)\},
\end{equation}
where $K_{\text{max}}$ is the maximum number of hypothesized exponentials (naturally, $K_{\text{max}}\leq L)$, $\mathbf{y} = \{y(n)\}_{n=0}^{N-1}$ is the vector of observations \eqref{eq:noisy_signal_model}, and $p_k(\mathbf{y}|\hat {\mathbf{v}}_k)$ is the probability density function (PDF) of $\mathbf{y}$ under the estimated parameter vector $\hat {\mathbf{v}}_k$, assuming the model order is $k$. Here, the estimated parameter vector is $\hat{\mathbf{v}}_k = [\{\hat b_i\}_{i=1}^{k}, \{\hat \alpha_i\}_{i=1}^{k}, \{\hat \theta_i\}_{i=1}^{k}]$, obtained by the MP method.
The term $C(k, N)$ is a model-complexity penalty term, used as a regularization for maximum-likelihood based methods.
We emphasize that this is a coupled detection-estimation approach, which requires prior knowledge of the likelihood function.

Typically, $K\leq \widehat K$, and the matrix $\A$ has $\widehat K$ non-zero eigenvalues. These eigenvalues may include both perturbed signal
poles and extraneous eigenvalues that do not correspond to the signal poles~\cite{hua1991svd, sarkar1995using, adve1997extrapolation}.
These extraneous components must therefore be identified and discarded. This additional classification step is central to the present work and follows the approach introduced in~\cite{segman2026structure}. Unlike the conventional model order detection methods reviewed here, that aim to detect the model order prior to the estimation step, the proposed approach intentionally allows $K < \widehat K$ and subsequently classifies the computed eigenvalues \emph{a posteriori}, discarding the extraneous ones. The proposed classification framework is developed in detail in the following sections.

\section{True and Spurious Components} \label{sec: component definition}

In this section, we define the notion of a computed eigenvalue being true or spurious, where a true eigenvalue is required to lie sufficiently close to at least one signal pole. By retaining only the components classified as true and using them for parameter estimation, we obtain theoretical guarantees on the resulting estimation accuracy under noise.

In the noiseless setting, with $\widehat K=K$, the computed eigenvalues in ~\eqref{eq:reduced_matrix_eigenvalue_eq} coincide with the $K$ distinct signal poles, establishing a one-to-one
association~\cite{hua1990matrix, hua1991svd, sarkar1995using, potts2013parameter}. 
In the presence of noise, however, the computed eigenvalues are perturbed and no longer coincide exactly with the signal poles. 
Even when the model order is known and $\widehat K = K$, the Cramér–Rao lower bound (CRB)~\cite{stoica2005spectral} implies a nonzero lower bound on the variance of any unbiased estimator, so exact pole recovery cannot generally be expected.
When $\widehat K>K$, the number of eigenvalues exceeds the number of signal poles, and include both
perturbed signal poles and extraneous eigenvalues that do not correspond to the signal poles.
Consequently, whether a computed eigenvalue should be regarded as \emph{true} (signal-related), or \emph{spurious} (noise-related) must be defined carefully.

We now formalize this distinction.
For each true pole $z_k$, we associate a prescribed tolerance $\Delta_k>0$ and define its acceptance disk by
\begin{equation}\label{eq:Delta_k_def}
    \cD_k := 
    \left\{\lambda\in\mathbb C:
    |\lambda-z_k|\leq\Delta_k\right\}.
\end{equation}
The overall acceptance region is then defined by
\begin{equation}\label{eq:acceptance region_D}
    \cD := \bigcup_{k=1}^K \cD_k.
\end{equation}
The region $\cD$ therefore defines a model-dependent acceptance region for the computed eigenvalues with bounded estimation error. This leads to the following definition.

\begin{definition}[True and spurious eigenvalues]
\label{def:true_spurious}
    Let $\lambda_i$ be a computed eigenvalue of $\A$ in~\eqref{eq:reduced_matrix_eigenvalue_eq}. Then $\lambda_i$ is called {true} (signal-related) if $\lambda_i\in\cD$. Otherwise, $\lambda_i$ is called {spurious} (noise-related).
\end{definition}
\noindent Equivalently, this can be expressed as
\begin{equation}
    \lambda_i \text{ is true } \iff \lambda_i\in\cD \iff d_{\cD}(\lambda_i)\leq1,
\end{equation}
where
\begin{equation}\label{eq:lambda_dist_to_D}
    d_{\cD}(\lambda) :=\min_{1\leq k\leq K} \frac{|\lambda-z_k|}{{\Delta_k}}.
\end{equation}

Definition~\ref{def:true_spurious} formalizes the model order detection problem as a classification of the computed eigenvalues. Accordingly, a true detection occurs when an eigenvalue in $\mathcal D$ is correctly classified as true, whereas a missed detection occurs when such an eigenvalue is discarded.

\section{The Proposed Score \& Algorithm} \label{sec: proposed score}
In this section, we propose an algorithm for classifying computed eigenvalues as true or spurious, using a new score computed for each spectral component.
The score measures the alignment of each component with the dominant right subspace of the observed data. We then establish a theoretical result showing that a large score guarantees that the associated eigenvalue is true according to Definition~\ref{def:true_spurious}. This links the classification of a true eigenvalue to its estimation accuracy, providing a detection step with estimation guarantees.

\subsection{The Proposed Score}


Let $(\lambda_i, \q_i)$ be a computed normalized eigenpair of $\A$ as defined in~\eqref{eq:reduced_matrix_eigenvalue_eq}, and denote the lifting of $\q_i$ to the observed data space by
\begin{equation}
    \mathbf d_i := \V_{\widehat K}\q_i \in\mathbb{C}^L.
    \label{eq:lifting_q}
\end{equation}
We refer to $\mathbf d_i$ as the \emph{right MP direction} since it satisfies the following projected MP equation 
\begin{equation}\label{eq:projected_MP_equation}
    \mathbf U_{\widehat K}^{\mathrm H}
    (\mathbf Y_1-\lambda_i\mathbf Y_0)
    \mathbf d_i =\mathbf 0,
\end{equation}
derived from the projected problem in~\eqref{eq:projected_problem}. 

The lifting of $\mathbf q_i$ \eqref{eq:lifting_q} allows us to compare it with the signal subspace in the data space.
In the noiseless case, with $\widehat K=K$, every right MP direction lies entirely in the clean right signal subspace $\Range(\X_0^{\mathrm H})$~\footnote{In the noiseless rank-$K$ case, the right singular vectors
corresponding to the nonzero singular values of $\X_0$ span
$\Range(\X_0^{\mathrm H})$. Since $\mathbf d_i$ is a linear
combination of these vectors, it belongs to this subspace.}.
In the presence of noise, since the clean right signal subspace has dimension $K$, when $\widehat K\geq K$ and the noise perturbation is sufficiently small, this subspace is approximated by the leading $K$ right singular vectors of $\Y_0$. Since $K$ is unknown, the alignment of $\mathbf d_i$ with the signal subspace is estimated by projecting it onto the dominant right singular subspace of $\Y_0$.

To this end, we introduce a score rank $1\leq p< \widehat K$, and project $\mathbf d_i$ onto the subspace spanned by the leading $p$ right singular vectors of $\Y_0$. Let $\V_p$ contain the $p$ leading right singular vectors of $\Y_0$, and define the corresponding orthogonal projector
\begin{equation}
    \mathbf P_{y,p} = \V_p\V_p^{\mathrm{H}}\in\mathbb{C}^{L\times L}.
    \label{eq:observed_projector}
\end{equation}
The range of $\mathbf P_{y,p}$ is the dominant $p$-dimensional right singular subspace of the observed data. Projecting $\mathbf d_i$ onto this subspace therefore measures how much of this direction lies in the leading $p$ right singular vectors. Accordingly, we define the following score.

\begin{definition}[The proposed score]
The rank-$p$ score of a computed eigenpair $(\lambda_i,\mathbf q_i)$ of $\A$ is defined by
\begin{equation}
    s_p(i) := \|\mathbf P_{y,p}\mathbf d_i\|^2.
    \label{eq:rank_p_score}
\end{equation}
\end{definition}
Since $\mathbf d_i=\V_{\widehat K}\q_i$ and the columns of
$\V_{\widehat K}$ are orthonormal, the score can be recast as
\begin{equation}
    s_p(i)=\sum_{\ell=1}^{p}|q_i(\ell)|^2,
    \label{eq:rank_p_score_recast_concentration}
\end{equation}
which is the energy concentrated at the first $p$ coordinates of the $i$th eigenvector $\mathbf q_i$. Since $\|\q_i\|=1$, the score is bounded $0\leq s_p(i)\leq1$, and a score close to one indicates that $\mathbf d_i$ lies primarily in this $p$-dimensional subspace. 
We note that when $p=\widehat K$, 
the score $s_{\widehat K}(i)=\|\q_i\|_2=1$ for every component $1\leq i \leq \widehat K$, and therefore we only consider $1 \le p <\widehat K$.

The score in~\eqref{eq:rank_p_score_recast_concentration} is computed directly from the data-driven eigenvector $\mathbf q_i$ and does not require any prior knowledge of the signal parameters. In the sequel, we show both theoretically and empirically that large values of $s_p(i)$ indicate that the $i$th eigenvalue is true.


\subsection{The Proposed Algorithm}
We now present the REECAP (Right Eigenvector Energy Concentration for Assessing candidate Poles) algorithm, summarized in Algorithm~\ref{alg:proposed_algo}, for estimating the signal poles by classifying the computed eigenvalues as true or spurious. The algorithm forms the spectral components of $\A$ defined in~\eqref{eq:reduced_matrix_eigenvalue_eq}, and computes the score \(s_p(i)\) for each component. It then retains components whose scores exceed prescribed thresholds $\gamma_{i,p}$, when available; otherwise, it clusters the scores into two groups and retains the higher-mean group. Retained eigenvalues are classified as true, and the remaining as spurious. The model order is determined by the number of eigenvalues classified as true.

\begin{algorithm}[t]
\caption{Proposed REECAP Algorithm}\label{alg:proposed_algo}
\begin{algorithmic}[1]
\renewcommand{\algorithmicrequire}{\textbf{Hyperparams:}}
\REQUIRE pencil-parameter $L$, working-rank $\widehat K$, score-rank $1\leq p < \widehat K$, optional thresholds $\gamma_{i,p}$
\renewcommand{\algorithmicrequire}{\textbf{Input:}}
\REQUIRE noisy measurements $\{y(n)\}_{n=0}^{N-1}$. 
               
\ENSURE estimates of the signal poles $\widehat{\mathcal Z}$, detected model order $ K_{\text{det}}$

\STATE Form the pair $(\Y_0, \Y_1)$ with $L$ columns from the noisy measurements $y(n)$.
\STATE Compute the rank-$\widehat K$ SVD
       $\Y_0\approx
       \U_{\widehat K}\boldsymbol{\Sigma}_{\widehat K}
       \V_{\widehat K}^{\mathrm H}$.

\STATE Form
       $\A\ \leftarrow
       \boldsymbol{\Sigma}_{\widehat K}^{-1}
       \U_{\widehat K}^{\mathrm H}\Y_1\V_{\widehat K}$.
       
\STATE Compute normalized right eigenpairs $(\lambda_i,\q_i)$ of $\A$.

\STATE Compute $s_p(i)$ for all $i=1,\ldots,\widehat K$.
\IF{thresholds are provided}
    \STATE $\widehat{\mathcal Z}\leftarrow
    \{\lambda_i:s_p(i)>\gamma_{i,p}\}$.
\ELSE
    \STATE Cluster $\{s_p(i)\}_{i=1}^{\widehat K}$ into two groups; let $\mathcal S_+$ index the group with the larger mean score.
    \STATE $\widehat{\mathcal Z}\leftarrow
    \{\lambda_i:i\in\mathcal \mathcal S_+\}$.
\ENDIF
\STATE Set $K_{\text{det}}\leftarrow |\widehat{\mathcal Z}|$
\end{algorithmic}
\label{alg1}
\end{algorithm}

Given $N$ noisy measurements $\{y(n)\}_{n=0}^{N-1}$, steps $1-4$ of the proposed algorithm follow the standard MP method; the Hankel matrices $\Y_0,\ \Y_1$ are formed by~\eqref{eq:full noisy Hankel matrix}, using the hyperparameter $L$, and the spectral components of the projected matrix $\A$, defined in~\eqref{eq:projected_matrix_def}, are computed from the truncated SVD of $\Y_0$, using the hyperparameter $\widehat K$.
The core of the proposed algorithm lies in lines $5-12$, where the score $s_p(i)$ is computed for each spectral component $(\lambda_i, \mathbf q_i)$, using the hyperparameter $p$.
If thresholds are provided, components satisfying $s_p(i)>\gamma_{i,p}$ are retained; otherwise, the scores are clustered into two groups and the higher-mean group is retained. The corresponding eigenvalues are classified as true and form \(\widehat{\mathcal Z}\), the set of estimated signal poles. The model order is determined by the cardinality of $\widehat{\mathcal Z}$.

In the following section we show theoretically that sufficiently large values of $s_p(i)$ guarantee that $\lambda_i$ is a true eigenvalue according to Definition~\ref{def:true_spurious}. This motivates retaining the higher-mean cluster when thresholds are unavailable.

The proposed algorithm follows the same \emph{coupled} detection-estimation approach as SAMP~\cite{segman2026structure}, classifying computed eigenvalues using information obtained during the estimation stage. However, the present method further provides a theoretical guarantee: under a suitable rank condition, an eigenvalue whose score exceeds a theoretical threshold lies within a prescribed tolerance of at least one signal pole. This is fundamentally different from the conventional \emph{decoupled} detection-estimation approach used in the MP method, where the model order is determined before the MP eigenvalues are computed. In this sense, the proposed approach is closer in spirit to ITC methods, which similarly use the estimated MP eigenvalues to determine the model order. However, unlike the proposed method, ITC are derived from statistical models and therefore require prior knowledge of the underlying likelihood function.
The selection and practical implications of the required hyperparameters $L$, $\widehat K$, and $p$, and the optional thresholds $\gamma_{i,p}$, are discussed in Section~\ref{subsec:practical_considerations}.

\begin{remark}
The proposed score follows a subspace-based principle similar to that underlying MUSIC~\cite{schmidt1986multiple}. In MUSIC, the data covariance matrix is used to estimate the signal and noise subspaces, and candidate steering vectors are tested according to their alignment with these subspaces. At a true direction, the steering vector lies approximately in the signal subspace and is therefore nearly orthogonal to the noise subspace. Our method does not scan over candidate steering vectors. Instead, each computed MP component is evaluated through the alignment of its right MP direction with the dominant right singular subspace of the data, as defined in~\eqref{eq:rank_p_score}. For normalized eigenvectors, this alignment reduces to measuring the concentration of eigenvector energy in the dominant right singular directions. In this geometric sense, the proposed approach is closer in spirit to MUSIC than to SAMP, which classifies spectral components by exploiting structural information encoded in the MP modes~\cite{segman2026structure}.
\end{remark}
\vspace{-5mm}
\subsection{Theoretical Analysis}

In this section, we show that there exists a threshold such that any spectral component whose score exceeds it is guaranteed to have its associated eigenvalue in $\mathcal D$, and is therefore classified as true by Definition~\ref{def:true_spurious}. This, in turn, guarantees that this eigenvalue lies within the prescribed tolerance $\Delta_k$ of at least one signal pole $z_k$.
%
%
The analysis applies for every choice of $1\leq p<\widehat K$, without requiring prior knowledge of $K$.

Let $\mathbf P_x$ denote the orthogonal projector onto the clean right signal subspace $\operatorname{Range}(\X_0^{\mathrm H})$. Using the Vandermonde factorization $\X_0 = \Z_L\B\Z_R$ in~\eqref{eq:clean_vandermonde_factorization}, it holds that $\mathbf P_x=\Z_R^\dagger\Z_R$. Additionally, let
\begin{equation}  
    \mu_{\bDelta} := 
    \sigma_K\!\left(
        \U_{\widehat K}^{\mathrm{H}} \Z_L \B\bDelta
    \right)\sigma_K(\Z_R),
    \label{eq:projected_pencil_constant}
\end{equation}
where $\bDelta = \diag(\Delta_1,\ldots,\Delta_K)$, and $\{\Delta_k\}_{k=1}^K$ are the tolerances defined by~\eqref{eq:Delta_k_def}. Since the signal poles are assumed distinct and \(K\leq \min\{N-L,L\}\), both $\Z_L$ and $\Z_R$ have rank $K$. Moreover, $\widehat K\geq K$, so the $K$th singular value in~\eqref{eq:projected_pencil_constant} is well defined.
Additionally, since $\bDelta$ is diagonal, $\mu_{\bDelta}$ is nondecreasing in each $\Delta_k$. In particular, if all tolerances are scaled by a common factor \(\tau>0\), then
\[
    \mu_{\tau\bDelta}=\tau\mu_{\bDelta}.
\]
Hence, increasing or decreasing the tolerances proportionally increases or decreases $\mu_{\bDelta}$ by the same factor.

\begin{proposition} If $\mu_{\bDelta}>0$, then
\begin{equation}\label{eq:score_condition}
    s_p(i) > \Theta_{i,p}^2 \ \Rightarrow \ \lambda_i \in D,
\end{equation}
    where
    \begin{equation}\label{eq:Theta_def}
        \Theta_{i,p} := \|\V_p^{\mathrm{H}}(\I-\mathbf P_x)\mathbf d_i\| + \frac{\|\U_{\widehat K}^{\mathrm{H}}(\W_1- \lambda_i \W_0)\mathbf d_i\|}{\mu_{\boldsymbol{\Delta}}},
    \end{equation}
\label{prop:Score_implications}
\end{proposition}
\begin{proof}
    See Appendix~\ref{app:Score_implications_proof} in the supplementary materials (SM).
\end{proof}
\noindent We note that $\mu_{\bDelta} \geq 0$ by definition, and that $\mu_{\bDelta} > 0$ is equivalent to $\operatorname{Rank}(\U_{\widehat K}^{\mathrm H}\Z_L)=K$, ensuring that projection onto $\operatorname{Range}(\U_{\widehat K})$ does not annihilate any nonzero vector in the clean left signal subspace $\operatorname{Range}(\Z_L)$. See Appendix~\ref{app:Score_implications_proof} in the SM for the proof. Moreover, $\mu_{\bDelta}$ depends on the conditioning of the Vandermonde matrices $\Z_L$ and $\Z_R$, which is controlled by their dimensions and the locations of the signal poles~\cite{aubel2019vandermonde}; $\mu_{\bDelta}$ also depends on the amplitudes and the choice of $\widehat K$.

Using~\eqref{eq:rank_p_score_recast_concentration}, this proposition relates the energy concentration at the first $p$ coordinates of the data-driven eigenvector $\mathbf q_i$ to the classification in Definition~\ref{def:true_spurious}. Specifically,  
\[
    s_p(i) > \Theta_{i,p}^2 \ \Rightarrow \ |\lambda_i-z_k| \leq \Delta_k,
\]
 for at least one signal pole $z_k$.
Critically, Proposition~\ref{prop:Score_implications} is meaningful only when $\Theta_{i,p} < 1 $, since $0\leq s_p(i) \leq 1$ by~\eqref{eq:rank_p_score_recast_concentration}.
Since 
\[
    \X_0(\I-\mathbf P_x) = \mathbf 0, \quad
    \U_p^{\mathrm H}\Y_0 = \boldsymbol{\Sigma}_p \V_p^{\mathrm H},
\]
for $\|\V_p^{\mathrm H}(\I-\mathbf P_x)\mathbf d_i\|<1$, the condition $\Theta_{i,p}<1$ is equivalent to
\begin{equation}\label{eq:mu_lower_bound}
    \mu_{\boldsymbol{\Delta}}>\frac{\|\U_{\widehat K}^{\mathrm H}
       (\W_1-\lambda_i\W_0)\mathbf d_i\|}{1-\|\boldsymbol{\Sigma}_p^{-1}\U_p^{\mathrm H}\W_0(\I-\mathbf P_x)\mathbf d_i\|}.
\end{equation}
This condition relates the projected noise contributions on the right-hand side of~\eqref{eq:mu_lower_bound} to the prescribed tolerances through $\mu_{\bDelta}$.
It ensures that the threshold remains below the maximum attainable score, allowing Proposition~\ref{prop:Score_implications} to classify components as true for the prescribed tolerances.
For a given noise level, sample size, and signal parameters, this condition therefore determines how tightly the acceptance regions can be chosen while keeping Proposition~\ref{prop:Score_implications} useful for classification.
Informally, this indicates that stronger noise may require larger tolerances to keep $\Theta_{i,p}<1$, whereas lower noise may allow tighter tolerances to satisfy this condition. 


\begin{remark}\label{remark:components_indefinability}
Although $\lambda_i\in\cD$ implies that $\lambda_i$ is associated with at least one signal pole, it does not guarantee a unique association with that pole.
A unique association of $\lambda_i$ with a signal pole $z_k$ is possible only when 
\[
\lambda_i\in\mathcal D_k,\quad
\lambda_i\notin\mathcal D_{m}\ \forall m\neq k.
\]
This association is unique in both directions if, in addition, $\lambda_\ell\notin\mathcal D_k$ for all $\ell\neq i$.
Even under these conditions, however, not every signal pole is necessarily associated with a computed eigenvalue.
In this work, we focus only on assessing membership in $\mathcal D$, rather than resolving unique association with a particular signal pole. Such a unique association is left for future work.
\end{remark}

\begin{figure}
    \centering
    \includegraphics[width=1\linewidth]{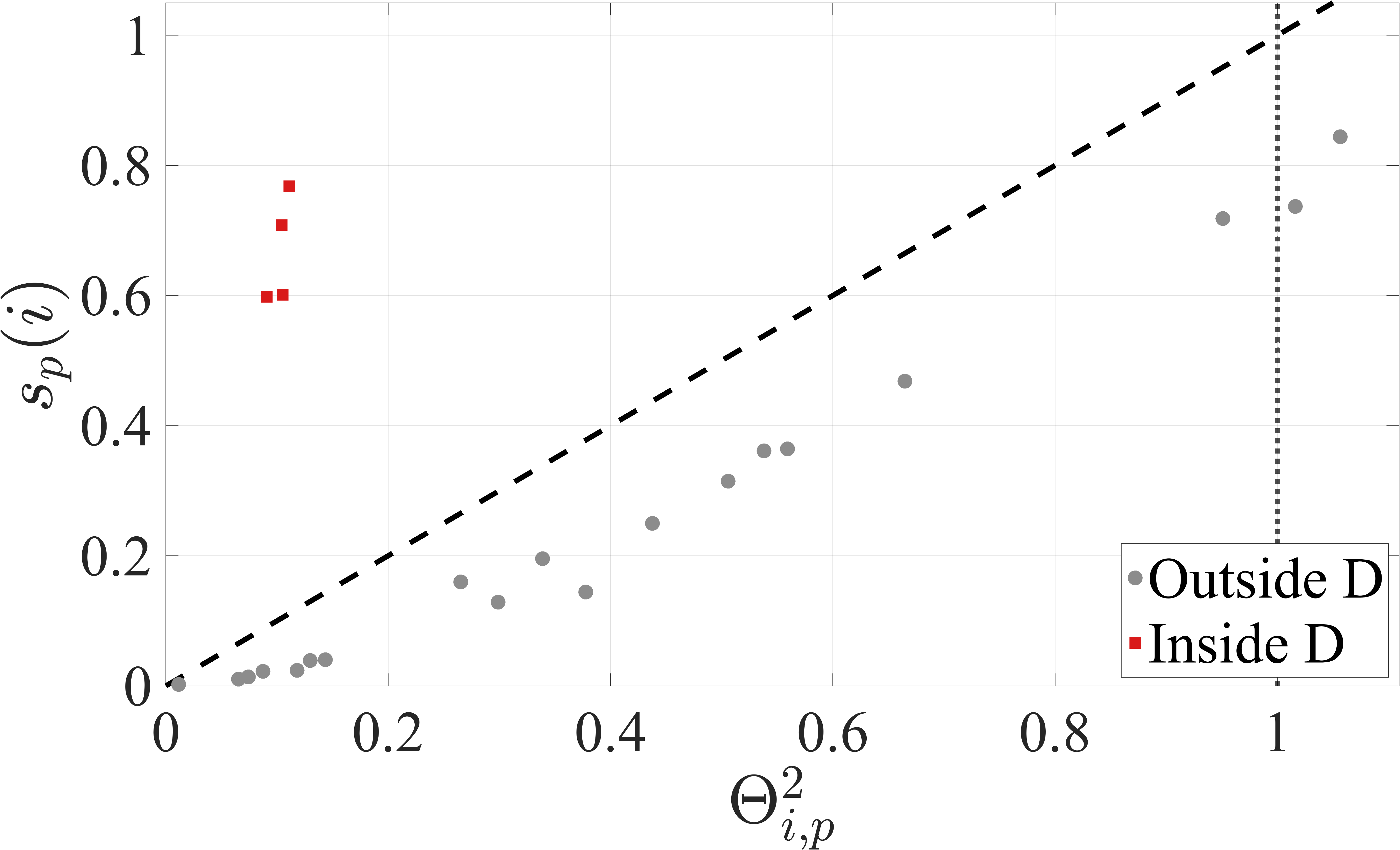}
\caption{Score $s_p(i)$ versus threshold $\Theta_{i,p}^2$ for a single noise realization. Red squares denote eigenvalues in $\mathcal D$; gray circles denote those outside. Points above the dashed diagonal satisfy the sufficient condition of Proposition~\ref{prop:Score_implications}. The dotted line marks $\Theta_{i,p}^2=1$.}
    \label{fig:proposition_1_illustration}
\end{figure}

\paragraph*{Illustrative example}
To illustrate a regime in which \(\Theta_{i,p}<1\), and the score identifies the true components in challenging conditions, consider the noisy signal model in~\eqref{eq:noisy_signal_model} with \(K=4\) damped exponentials partitioned into two clusters of equal size, where the angular frequency spacing within each cluster is the Rayleigh limit, and the clusters are separated by $15\times$ Rayleigh limit. The damping factors are drawn uniformly from \([0,0.05]\), while the complex amplitudes have magnitudes uniformly distributed over \([0.8,1.2]\) and phases uniformly distributed over \([0,2\pi)\). We use \(N=71\) samples, \(L=\lceil N/3\rceil\), and \(\widehat K=L-1\).
Figure~\ref{fig:proposition_1_illustration} shows $s_p(i)$ versus $\Theta_{i,p}^2$, computed according to~\eqref{eq:Theta_def}, for a single realization at an SNR of $10$ dB, and score-rank of $p=10$.
We set \(\Delta_k\) equals 2$\times$Rayleigh limit for all \(k\). Under the
stated damping range, membership in \(\mathcal D\) then guarantees an angular error bounded by approximately twice the Rayleigh limit, relative to at least one signal pole.
Condition~\eqref{eq:mu_lower_bound} holds for $21$ of the $23$ candidates; the remaining two candidates appear to the right of the dotted line at $\Theta_{i,p}^2=1$ and lie outside $\mathcal D$. 

The figure illustrates the guarantee of Proposition~\ref{prop:Score_implications}: components with
$s_p(i)>\Theta_{i,p}^2$, corresponding to points above the dashed diagonal, lie in $\mathcal D$. In this example, we also see empirically two important properties: (i) all four candidate eigenvalues in $\mathcal D$ satisfy condition~\eqref{eq:score_condition}, with no missed detections, and (ii) points classified as true lie well above the diagonal, while all
components outside $\mathcal D$ lie below it. Thus, all four components in $\mathcal D$ are accepted, with substantial margins above their thresholds, and with no false positives.

Finally, we remark that Proposition~\ref{prop:Score_implications} guarantees that components satisfying condition~\eqref{eq:score_condition} are true according to Definition~\ref{def:true_spurious}. 
However, this condition is only sufficient: components that fall below the threshold are not necessarily spurious. 
Thus, it does not guarantee that every true component satisfies condition~\eqref{eq:score_condition}, or that every signal pole is detected.
Moreover, although the score itself is computed from the observations, the theoretical threshold in~\eqref{eq:Theta_def} depends on unknown clean signal and noise quantities. In particular, the practical clustering-based classification rule in REECAP does not necessarily satisfy the sufficient condition of Proposition~\ref{prop:Score_implications}. It should therefore be interpreted as a theoretical guarantee rather than an implementable classification rule. 

\subsection{Computational Complexity }\label{susec:comput_complex}
We evaluate the computational complexity of Algorithm~\ref{alg:proposed_algo} by counting the number of complex operations each step requires, assuming that multiplying an $a\times b$ matrix by a $b\times c$ matrix requires $\mathcal O(abc)$ operations.
Computing an economy SVD of $\Y_0$ and retaining its leading $\widehat K$ components in step~$2$ requires $\mathcal O((N-L)L^2)$ operations.
Forming $\A$ in Step~$3$ requires $\mathcal O((N-L)L\widehat K+L\widehat K^2)$ operations, taking into account that $\boldsymbol{\Sigma}_{\widehat K}$ is diagonal. Computing all eigenpairs of $\A$ in step~$4$ requires $\mathcal O(\widehat K^3)$ operations.
Computing the scores $s_p(i)$ for all $\widehat K$ eigenvalues in step~$5$ requires $\mathcal O(p \widehat K)$ operations. If thresholds are provided, comparing the scores with their corresponding thresholds $\gamma_{i,p}$ and storing the accepted eigenvalues in $\widehat{\mathcal Z}$ requires $\mathcal O(\widehat K)$ operations. 
Otherwise, applying common two-cluster methods, such as $k$-means or GMM~\cite{qiu2007comparative}, to the $\widehat K$ scores typically requires $\mathcal O(RT\widehat K)$ operations, where $R$ and $T$ denote the numbers of restarts and iterations, respectively.

Hence, for fixed $R$ and $T$, since $p\leq\widehat K\leq L$ and $N/3\leq L\leq N/2$, a total of $\mathcal O(N^3)$ operations is required for the proposed Algorithm~\ref{alg:proposed_algo}. For comparison, the conventional MP algorithm, which estimates the model order by inspecting the singular values of $\Y_0$, has the same asymptotic complexity of $\mathcal O(N^3)$.

\subsection{Practical Considerations}\label{subsec:practical_considerations}

The proposed algorithm requires the hyperparameters $L$, $\widehat K$, $p$, and optionally $\gamma_{i,p}$ to be specified. The choice of the pencil parameter $L$ is well studied in the MP literature. As discussed in Section~\ref{subsec:background_MP}, it should satisfy $K \leq L \leq N-K$, with the practical range of $\frac{N}{3}\leq L\leq\frac{N}{2}$. Unless stated otherwise, we set $L=\lceil N/3\rceil$. The sensitivity of the proposed score to the choice of $L$ is examined in Section~\ref{subsec:results_sensitivity_analysis}.

The working rank \(1\leq \widehat K\leq L\), defined in~\eqref{eq:working_rank_def}, reflects any prior knowledge available about the number of signal poles $K$. A common example is an upper bound on $K$. For instance, if one assumes $2K\leq L$, then \(\widehat K=\lceil L/2\rceil\) may be selected. Choosing \(\widehat K=L\) corresponds to making no prior assumption on \(K\). The sensitivity of the proposed score to the choice of \(\widehat K\) is examined in Section~\ref{subsec:results_sensitivity_analysis}.

The score rank \(1\leq p<\widehat K\) determines the dimension of \(\V_p\) used to define \(s_p(i)\) in~\eqref{eq:rank_p_score}. Since the score estimates the alignment of each eigenvector with the dominant signal subspace, \(p\) should reflect the number of significant signal directions in \(\Y_0\), which may be smaller than the number of signal poles when poles are clustered.
In practice, $p$ can be estimated from the dominant singular values of $\Y_0$. See Section~\ref{sec:results} for more details and several heuristics for selecting $p$.

Finally, the thresholds $\gamma_{i,p}$ control the tradeoff between missed detections and false positives. In practice, they may be selected using an independent labeled calibration set, for example to achieve a target false-positive rate. Otherwise, if thresholds are unavailable, REECAP clusters the scores \(s_p(i)\) into two groups and retains the eigenvalues associated with the higher-mean cluster, without prescribing a threshold in advance.
We evaluate the score's ability to distinguish true from spurious components using the area under the ROC curve, obtained by sweeping the threshold over \([0,1]\). See Section~\ref{sec:results} for more details.

\subsection{Relation to Other Methods}\label{subsec:prior_works}

Several prior works have considered similar \emph{a posteriori} approaches to detecting the model order.
Within MP, SAMP~\cite{segman2026structure} introduces the MP-modes, defined as the pseudo-inverse matrix of the generalized eigenvectors matrix obtained by the MP method. It evaluates the computed eigenvalues using the geometric structure imposed on these modes by the exponential signal model. SAMP++ additionally exploits dependencies among the modes~\cite{segman2025samp++}. Both methods rely on maximization over a dense grid, leading to computational cost that increases with the grid resolution. Moreover, neither method provides estimation-accuracy guarantees for the accepted eigenvalues.

Within ESPRIT, ESTER links model order detection to estimation-error bounds, using a shift-invariance residual, derived only for model order underestimation in a noiseless setting. Its direct implementation is computationally demanding; a recursive version reduces this cost but still requires auxiliary matrix updates for candidate model orders~\cite{badeau2006new}. 
%
RIVES applies ESPRIT with an overestimated order and screens the resulting estimates using angles between reconstructed exponential vectors and an estimated noise subspace, without guaranteeing estimation-accuracy for the accepted estimates~\cite{yildirim2022multi}. 

In contrast, our analysis explicitly accounts for additive observation noise, provides a sufficient condition for accepting individual estimates within a prescribed estimation tolerance, and employs a simple and efficient score based on the energy concentration of the already computed eigenvectors, without requiring auxiliary matrix construction or grid search.

\section{Numerical Results} \label{sec:results}
In this section, we evaluate the performance of the proposed method through extensive Monte Carlo simulations across different SNR regimes and signal configurations.
We compare REECAP (Algorithm~\ref{alg:proposed_algo}) with several representative baselines. These include the recently introduced SAMP methods~\cite{segman2026structure,segman2025samp++}; conventional model order detection techniques used with the MP method, SDD~\cite{sarkar1995using} and spectral-gap~\cite{yin2011model} criteria, reviewed in Section~\ref{subsec:MP_model_order_detection}; the two related methods ESTER~\cite{badeau2006new} and RIVES~\cite{yildirim2022multi}, reviewed in Section~\ref{subsec:prior_works}; and two commonly used ITC-based methods, AIC~\cite{akaike1974new} and EVT~\cite{nadler2011model}.
The ITC-based methods use a coupled detection and estimation approach that requires prior knowledge of the noise distribution (that is unknown in our setting) and involves computing the likelihood function for every possible hypothesized model order, making them computationally inefficient. The EVT criterion was shown to outperform the common MDL and MAP criteria, and to be less sensitive to noise model mismatch~\cite{nadler2011model}. The code to reproduce the numerical results in this section is available in the following \href{https://github.com/dsadka/REECAP--Right-Eigenvector-Energy-Concentration-Assessing-Poles}{GitHub} link.

The proposed REECAP is based on the score $s_p(i)$, which depends on the hyperparameter $1\leq p <\widehat K$. We select $p$ by estimating the number of dominant singular values of $\Y_0$, using the elbow method~\cite{cattell1966scree}, which identifies the transition from a steep decline to a flatter tail in the ordered singular-value curve. The sensitivity of REECAP to the choice of \(p\) is examined in Section~\ref{subsec:results_sensitivity_analysis}, and several heuristics for selecting $p$ are compared in Appendix~\ref{app:heuristics_for_selecting_p} of the SM.

\subsection{Simulation Setup}

We consider the signal model in~\eqref{eq:noisy_signal_model} with zero-mean, temporally white Gaussian noise. The $K$ signal poles form clusters in the complex plane, following the clustered source configurations studied in super-resolution limits of subspace-based methods~\cite{li2020super,batenkov2021super, diab2025selecting}. 
We control the number of clusters, the number of poles within each cluster, and the minimum pairwise pole distance taken over the entire pole set. We note that theoretical frequency resolution limits are also discussed in~\cite{shahram2005resolvability, amar2008fundamental}. 

Unless it is the swept variable or otherwise stated, we use $N=71$, $L=\lceil N/3\rceil$, $\widehat K=L-1$, and $K=6$ poles arranged in three clusters, each containing two poles, with a within-cluster frequency spacing of one Rayleigh limit and an inter-cluster frequency spacing, measured between cluster centers, of $15$ Rayleigh limits, following~\cite{li2020super}. We set the pole amplitudes to $b_k=1$ and draw the damping factors uniformly from $[0,0.05]$. The default SNR is $0$~dB, with
\[
    \mathrm{SNR}
    =10\log_{10}\!\left(\frac{\sum_{k=1}^{K}|b_k|^2}{\sigma_w^2}\right),
\]
where $\sigma_w^2$ is the noise variance. Results are averaged over $500$ independent Monte Carlo trials.

For methods with a decision threshold, we report the area under the receiver operating characteristic (ROC) curve (AUC-ROC). 
The ROC curve is obtained by varying the threshold and plotting the
fraction of true eigenvalues correctly accepted (TPR) against the fraction of spurious eigenvalues incorrectly accepted (FPR). The AUC-ROC therefore summarizes the eigenvalues classification performance over all threshold values, with larger values indicating better separation between true and spurious eigenvalues.
Ground-truth labels are determined by whether the computed eigenvalues lie within the acceptance region \(\mathcal D\), defined in~\eqref{eq:acceptance region_D}, using a common tolerance of $\Delta_k=2\times$Rayleigh limit.
This evaluation couples detection with an estimation-accuracy requirement: an accepted component counts as a true positive only if its eigenvalue lies within the prescribed tolerance $\Delta_k$ of at least one signal pole. 
For comparison with threshold-free ITC-based methods, we also report the probability of correct model-order detection, $P_d = \mathbb{P}( K_{\text{det}} = K)$, estimated as the fraction of trials in which the detected model order $K_{\mathrm{det}}$ equals $K$.

Assuming a known model order, other geometrically motivated choices of $\Delta_k$ are possible. The deterministic bound~\cite{aubel2016deterministic}, $|\lambda_{\pi(k)}-z_k|\leq\eta_k$, where $\pi$ matches the $K$ estimates to the $K$ signal poles, motivates setting $\Delta_k=\eta_k$, ensuring that all $K$ estimates lie in \(\mathcal D\) under the bound's assumptions. Alternatively, the first-order pole perturbation~\cite{hua1990matrix}
\[
    \delta z_k =
    \frac{\boldsymbol l_k^{\mathrm H}(\W_1-z_k\W_0)\boldsymbol r_k}{b_k},
\]
where $\boldsymbol l_k$ and $\boldsymbol r_k$ are the left and right eigenvectors of the noiseless pencil defined in~\eqref{eq:clean_pencil_factorization}, motivates setting $\Delta_k=c_k|\delta z_k|$, with $c_k\geq1$. 
Both choices relate the tolerances $\Delta_k$ to pole-estimation errors, but assume known model order and depend on quantities unavailable in practice.
In the default configuration, $\Delta_k=c_k|\delta z_k|$ with $c_k=2$ yields tolerances comparable to $\Delta_k=2\times$ the Rayleigh limit used in our simulations, providing additional motivation for our choice.

\subsection{Comparison with Baselines}\label{subsec:results_baselines_comparison}

\begin{figure}[t]
    \centering
    \includegraphics[width=0.88\linewidth]{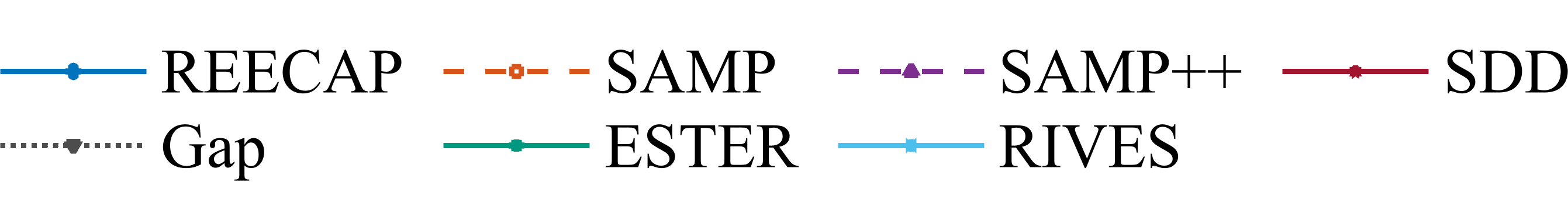}

    \begin{subfigure}[t]{0.48\linewidth}
        \centering
        \includegraphics[width=\linewidth]{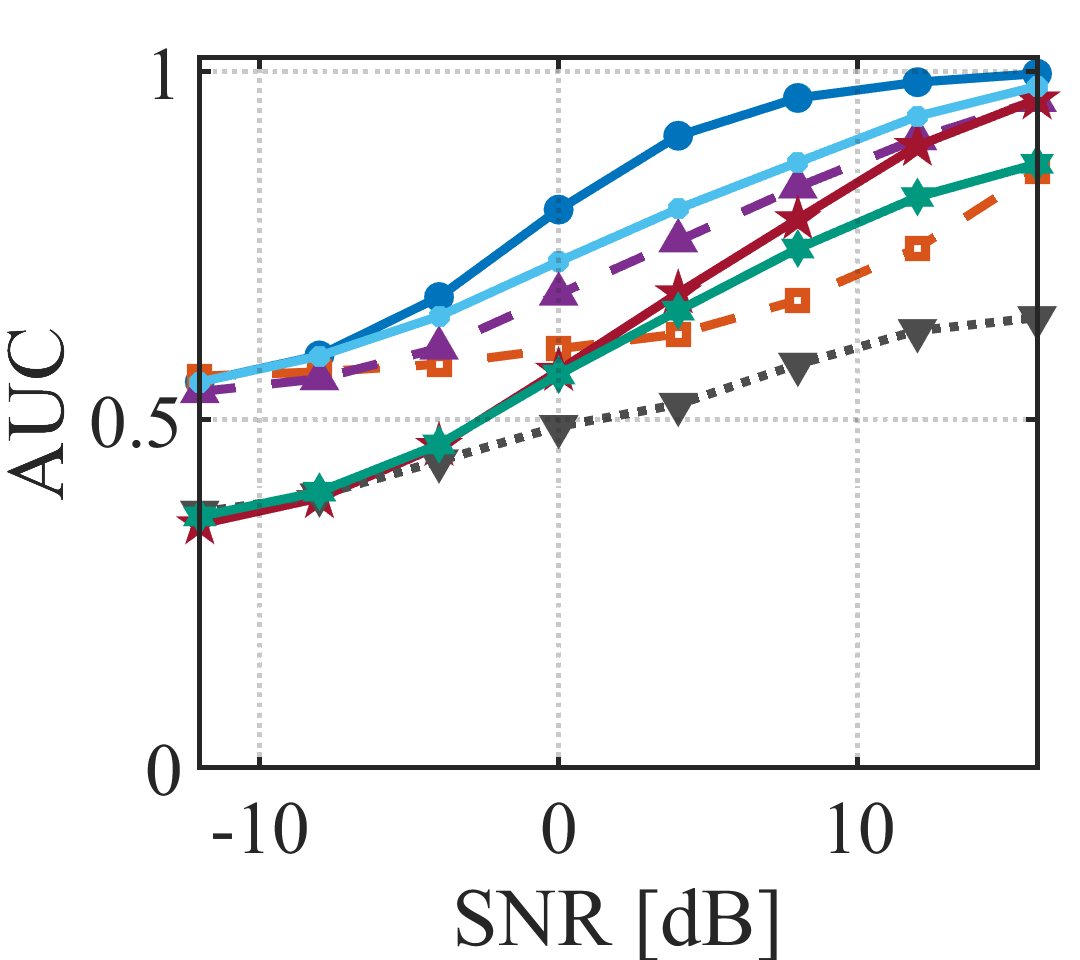}
        \caption{}
        \label{fig:AUC_SNR}
    \end{subfigure}\hfill
    \begin{subfigure}[t]{0.48\linewidth}
        \centering
        \includegraphics[width=\linewidth]{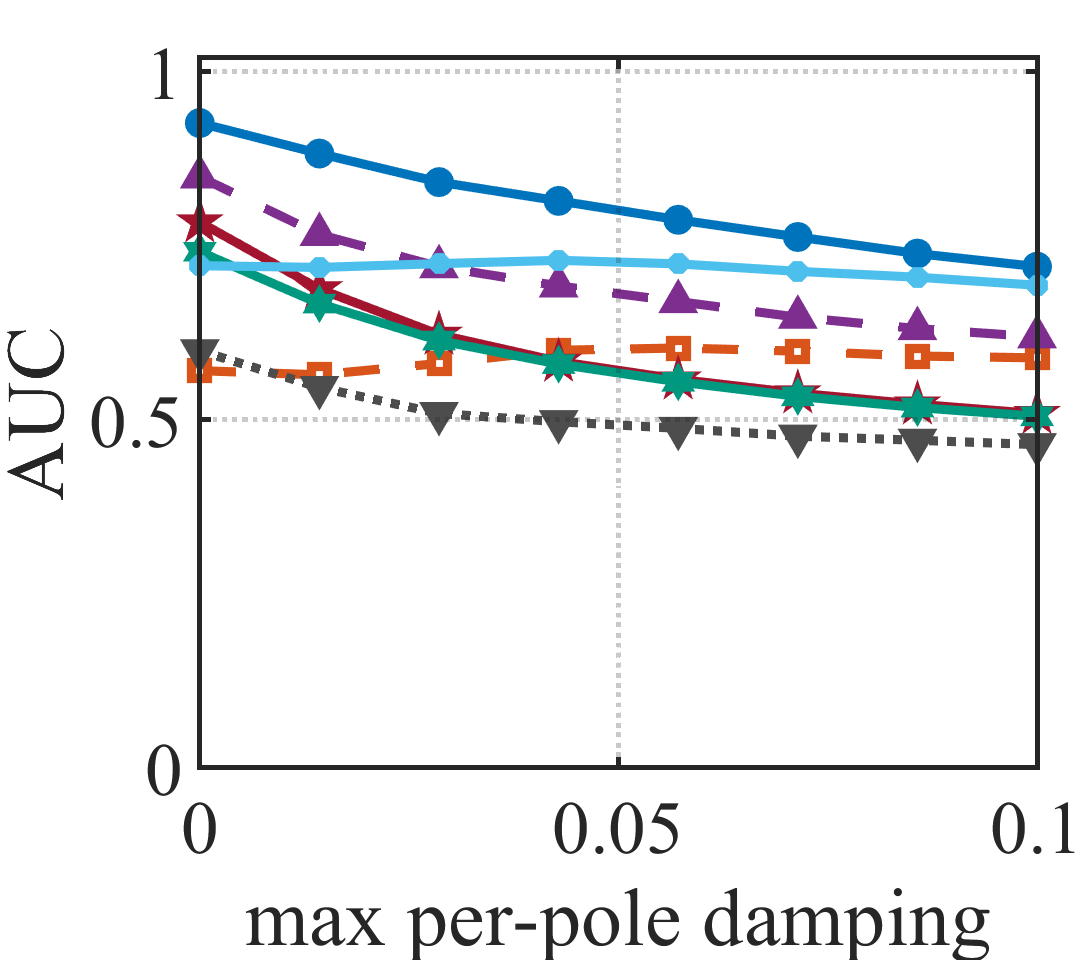}
        \caption{}
        \label{fig:AUC_damping}
    \end{subfigure}


    \begin{subfigure}[t]{0.48\linewidth}
        \centering
        \includegraphics[width=\linewidth]{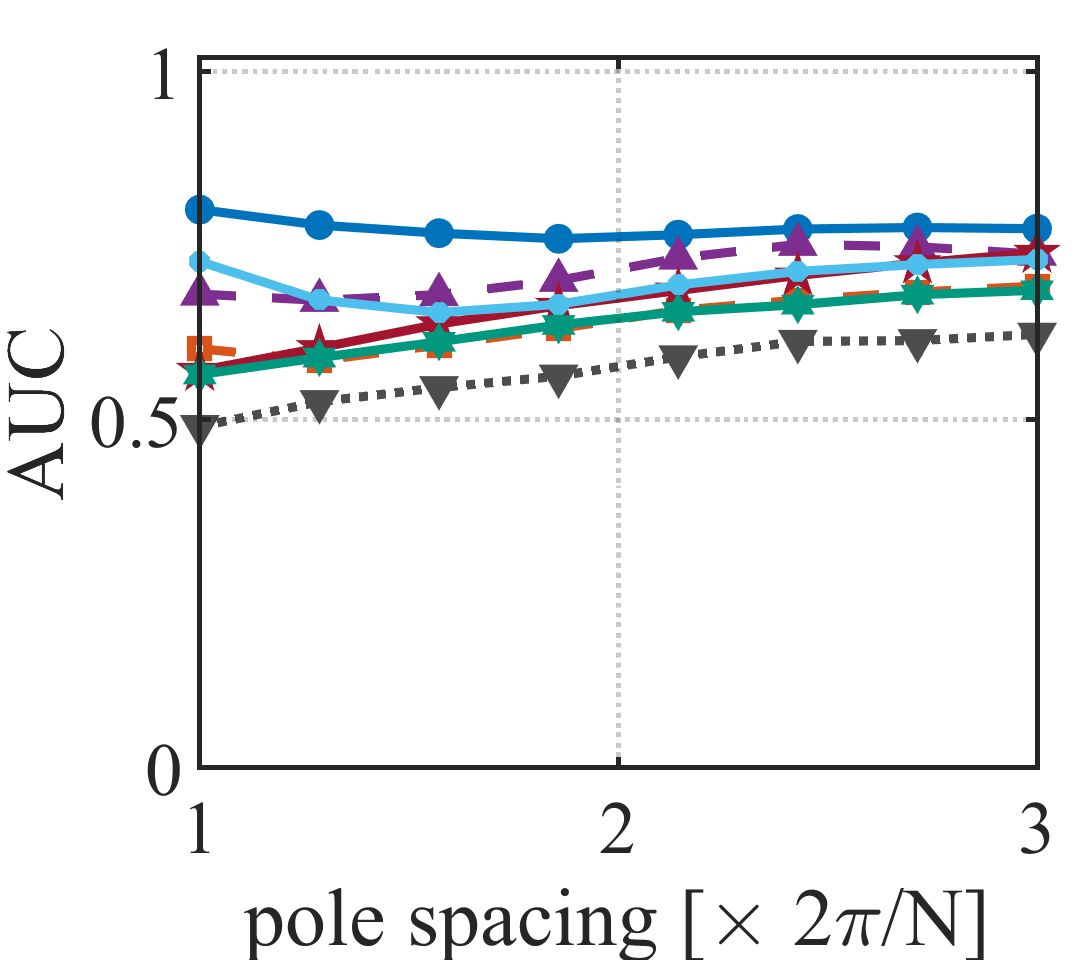}
        \caption{}
        \label{fig:AUC_pole_spacing}
        \label{fig:auc_K}
    \end{subfigure}\hfill
    \begin{subfigure}[t]{0.48\linewidth}
        \centering
        \includegraphics[width=\linewidth]{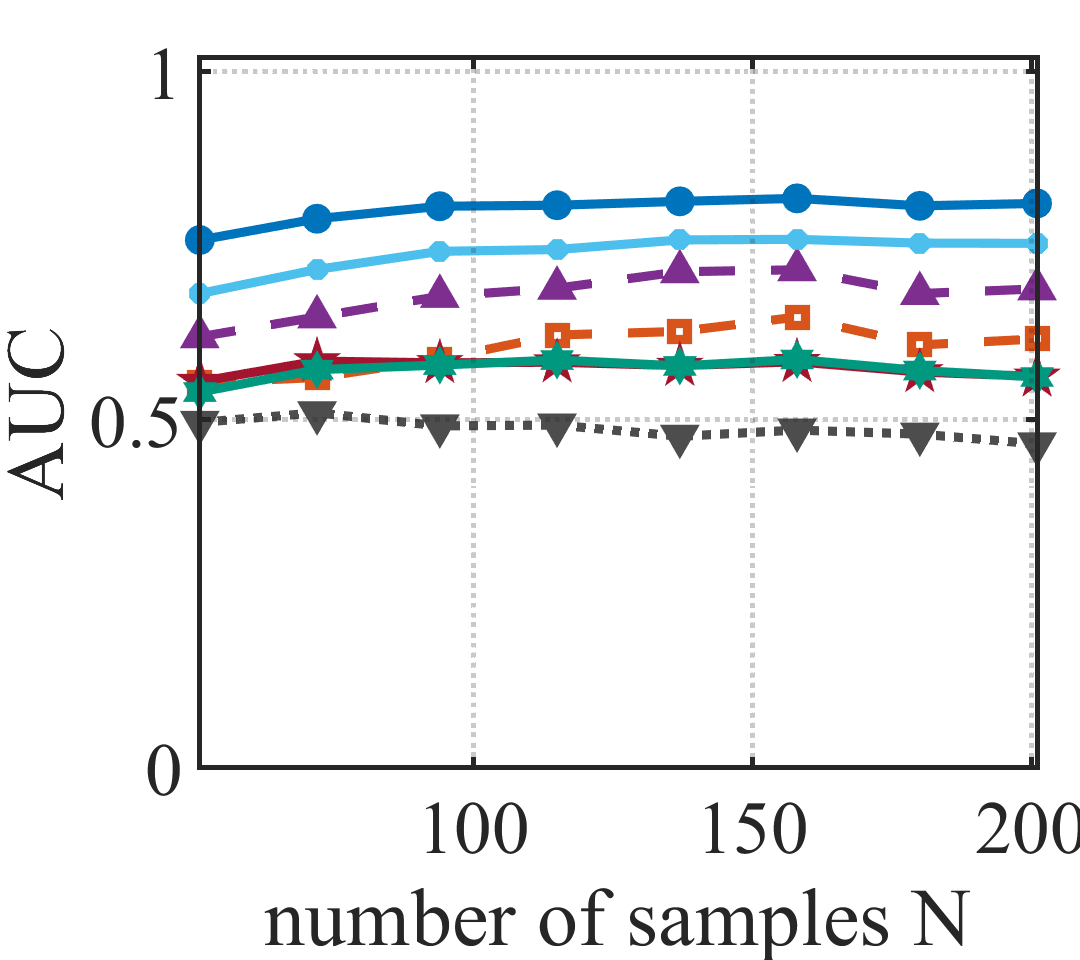}
        \caption{}
        \label{fig:AUC_N}
    \end{subfigure}


    \begin{subfigure}[t]{0.48\linewidth}
        \centering
        \includegraphics[width=\linewidth]{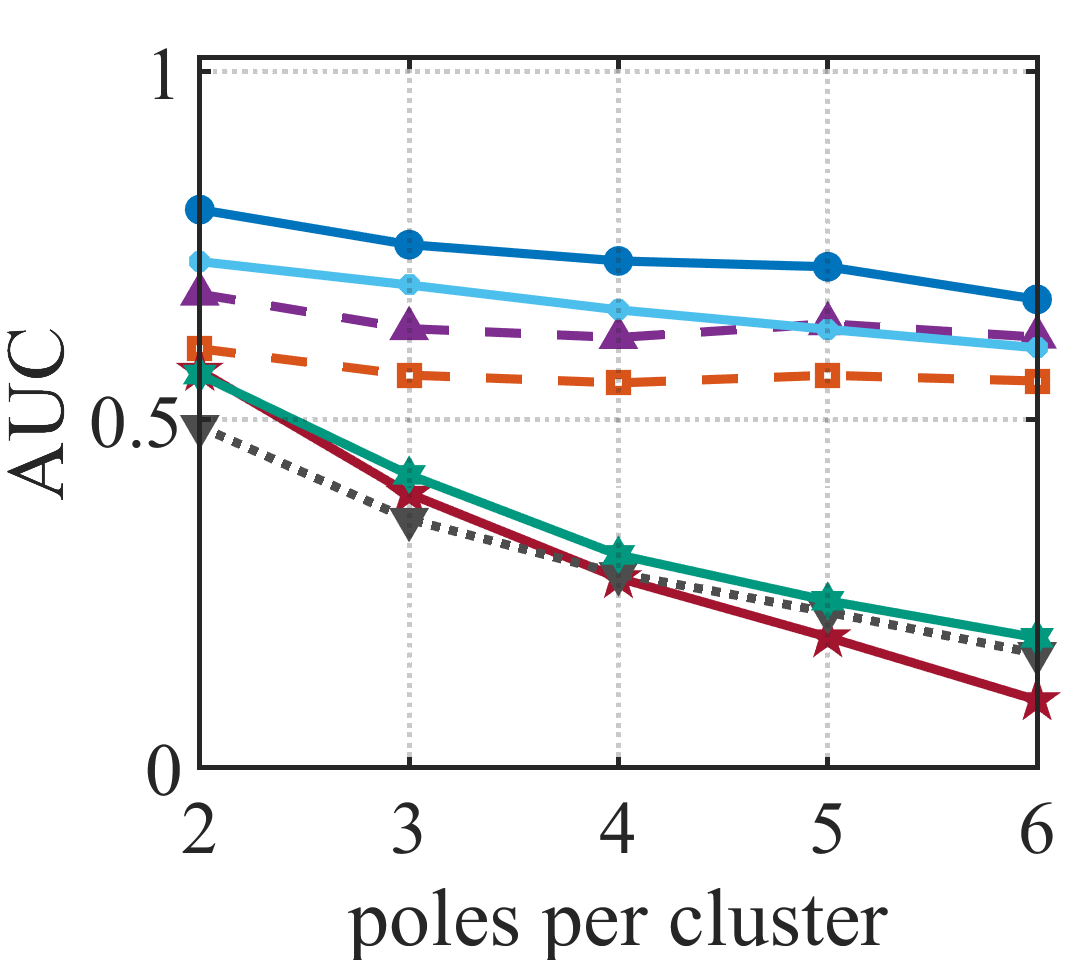}
        \caption{}
        \label{fig:AUC_poles_per_clust}
    \end{subfigure}\hfill
    \begin{subfigure}[t]{0.48\linewidth}
        \centering
        \includegraphics[width=\linewidth]{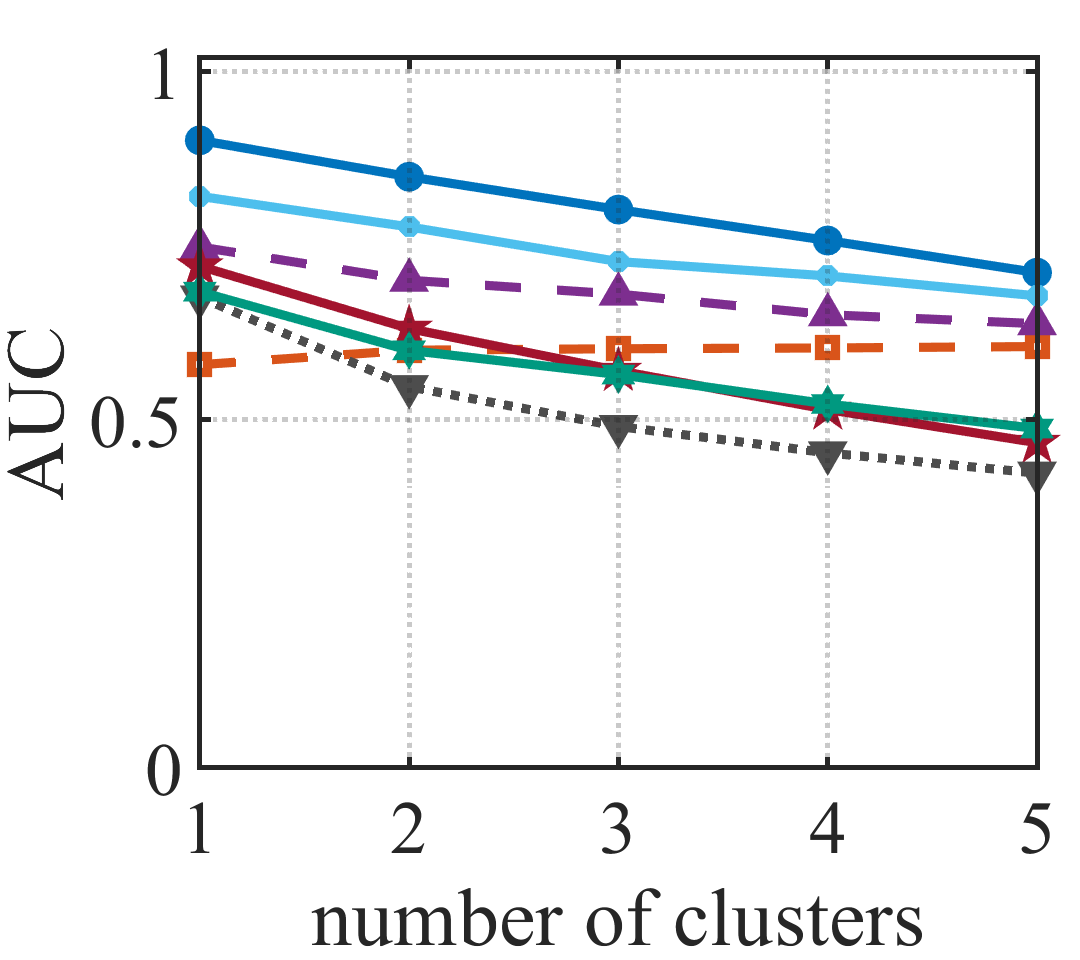}
        \caption{}
        \label{fig:AUC_nClust}
    \end{subfigure}
    \caption{AUC-ROC comparison of the proposed REECAP algorithm and the baseline methods across the simulation parameters.}
    \label{fig:AUC_baseline_compare}
\end{figure}

Fig.~\ref{fig:AUC_baseline_compare} compares the AUC-ROC of the proposed method and the threshold-based competing methods across a broad range of simulation parameters.
The proposed method generally achieves the highest AUC across all tested scenarios, with a notable margin except for the lowest tested SNR. 
Specifically, its AUC approaches one as SNR increases, with the largest gains over the baselines at intermediate SNRs. 
Increasing the sample size initially improve its AUC, which then stays constant around $0.8$. 
Increasing damping, the number of poles per cluster, or the number of clusters lowers the proposed AUC, but it remains higher than the baselines.
Notably, SDD, ESTER, and Gap methods deteriorate sharply as the number of poles per cluster increases.
Larger pole separations generally improve most baselines, reducing the advantage of the proposed method over the baselines.

\begin{figure}[t]
    \centering

    \includegraphics[width=0.96\linewidth]{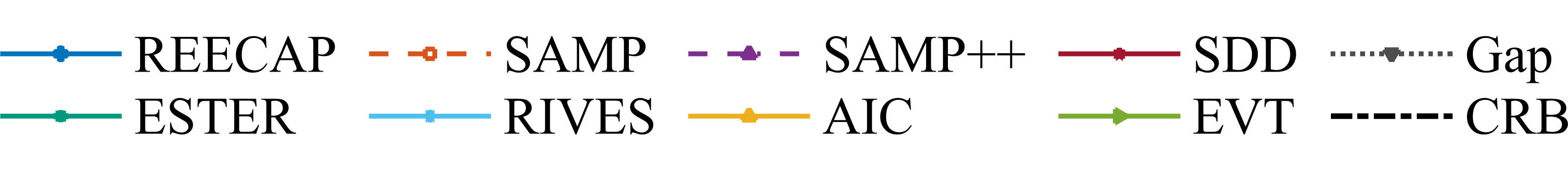}

    \begin{subfigure}[t]{0.48\linewidth}
        \centering
        \includegraphics[width=\linewidth]{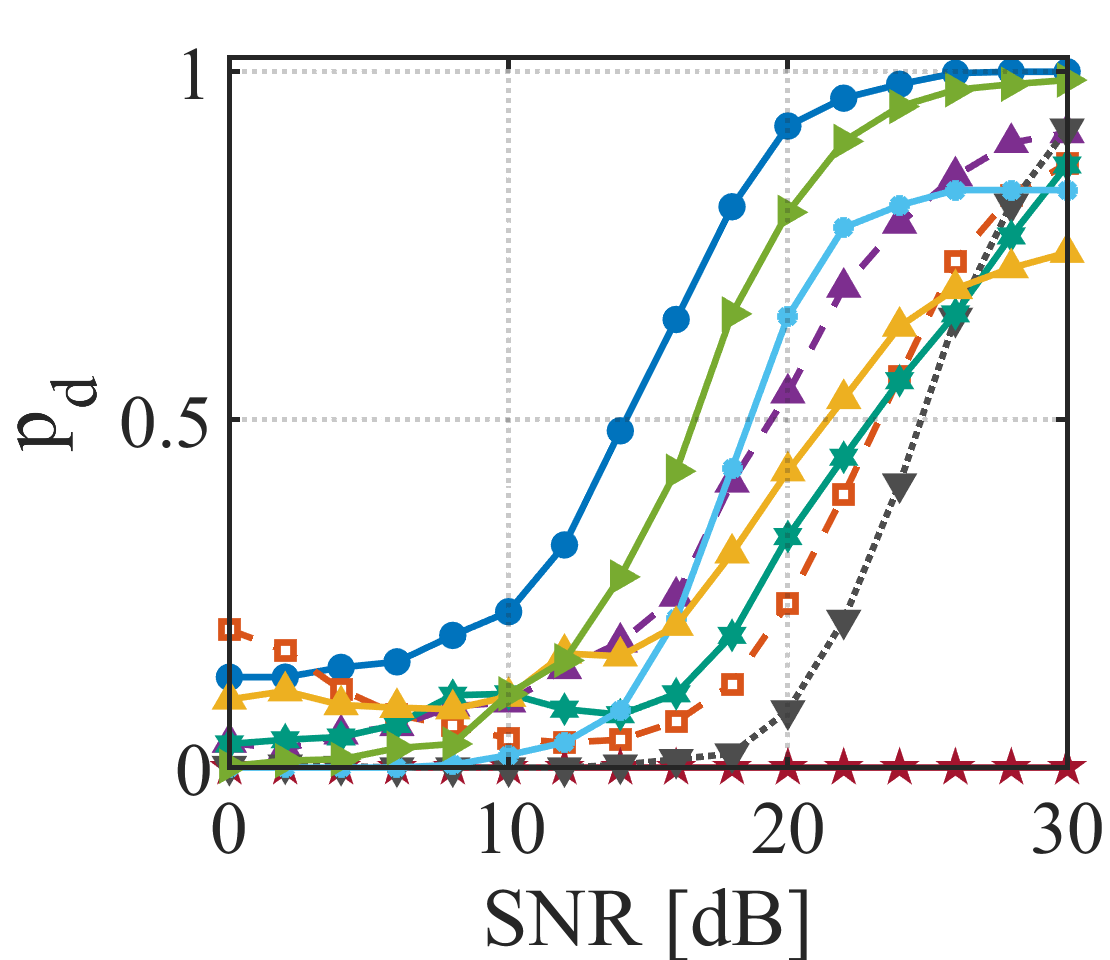}
        \caption{}
        \label{fig:pd_vs_snr}
    \end{subfigure}\hfill
    \begin{subfigure}[t]{0.48\linewidth}
        \centering
        \includegraphics[width=\linewidth]{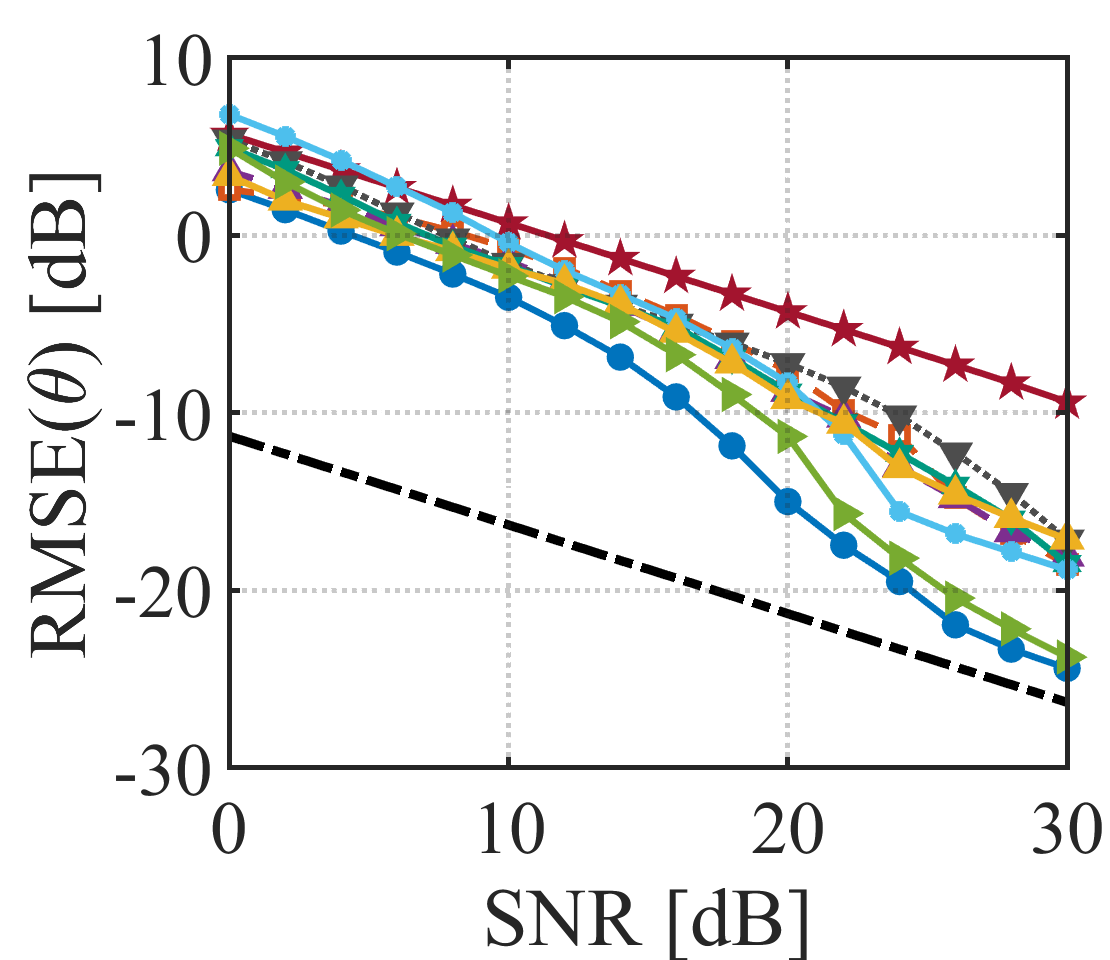}
        \caption{}
        \label{fig:RMSE_vs_snr}
    \end{subfigure}

    \begin{subfigure}[t]{0.48\linewidth}
        \centering
        \includegraphics[width=\linewidth]{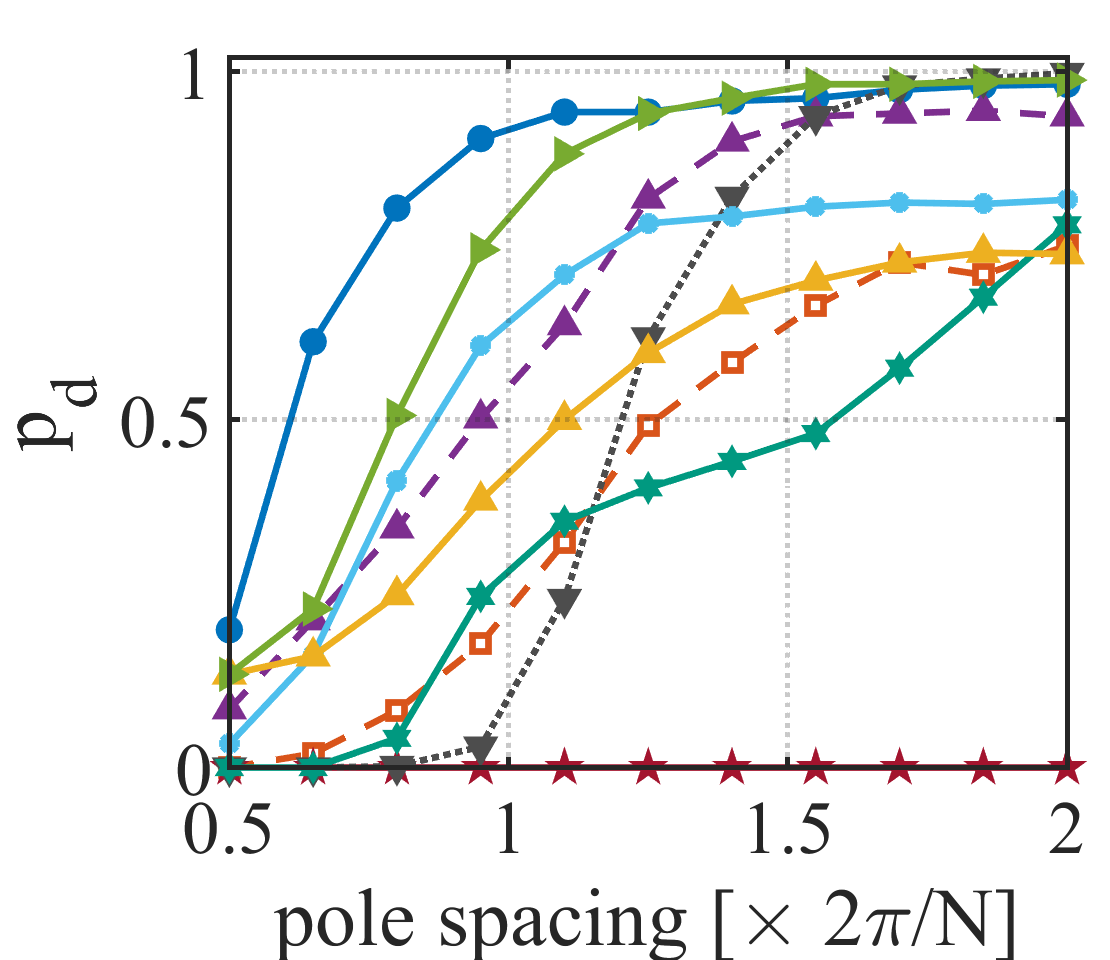}
        \caption{}
        \label{fig:pd_vs_pole_spacing}
    \end{subfigure}\hfill
    \begin{subfigure}[t]{0.48\linewidth}
        \centering
        \includegraphics[width=\linewidth]{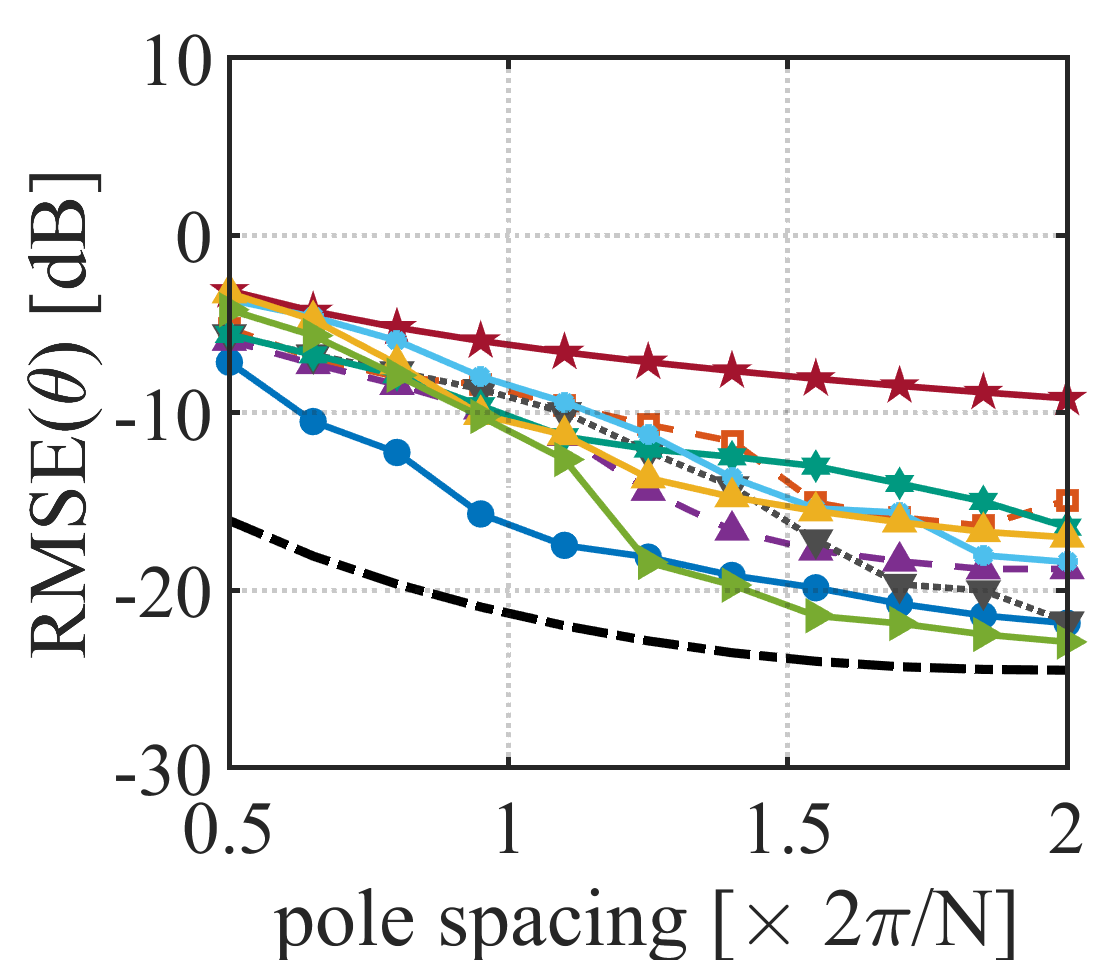}
        \caption{}
        \label{fig:RMSE_vs_pole_spacing}
    \end{subfigure}
    
    \caption{Probability of correct model order detection (a), (c) and average frequency-estimation RMSE (b), (d) versus SNR (top) and within-cluster pole spacing (bottom). The pole spacing is fixed at one Rayleigh limit in (a), (b), while the SNR is fixed at \(20\) dB in (c), (d).}
    \label{fig:pd_and_RMSE_vs_snr}
\end{figure}

Fig.~\ref{fig:pd_and_RMSE_vs_snr} compares $P_d$ and the average RMSE of the estimates $\{\hat \theta_k\}_{k=1}^K$ of the proposed method, the threshold-based methods, and the ITC-based methods versus SNR and the minimal pole distance within clusters. 
Candidate eigenvalues are matched to the nearest signal poles.
For trials with incorrect model-order detection, we assign a penalty proportional to the model-order error and scaled by the evaluation variable. For example, when evaluating RMSE versus pole separation $d_z$, the penalty is $|K_{\text{det}}-K|/d_z$. This provides consistent penalization across parameter ranges.

We include the CRB as a reference for evaluating the RMSE of the estimates $\{\hat \theta_k\}_{k=1}^K$~\cite{stoica2005spectral,lang1980frequency}.
We note that the CRB is a local bound and becomes tight only asymptotically. Global alternatives, such as the recently proposed Ziv–Zakai bounds for multiple-source parameter estimation \cite{zhang2022ziv}, have been shown to remain tight over a broad range of SNR values.
Since we do not assume the availability of a calibration set for tuning the decision thresholds, REECAP uses \(k\)-means clustering applied on the log scores. For RIVES, which does not prescribe a specific threshold, we use a numerically optimized value. For SAMP++, we use \(k\)-means clustering as recommended in the original paper, while for the remaining threshold-based methods we use the thresholds recommended in their respective publications.

The proposed method achieves the highest \(P_d\) and the lowest average RMSE across the tested SNR range, except at the two lowest SNR points. At high SNR values, its detection probability approaches one, and its RMSE approaches the CRB. Although RIVES was the strongest competitor to REECAP in Fig.~\ref{fig:AUC_baseline_compare}, even with a threshold optimized for this simulation its performance is substantially lower than that of REECAP, and EVT becomes the closest competing method.
Particularly, at large pole spacing above $1.25\times$Rayleigh limit, EVT achieves lower RMSE than REECAP.

\textit{Noise distribution effects:} To assess the impact of deviations from the Gaussian noise assumption, we also simulate non-Gaussian additive noise. Table \ref{table:pd_auc_noise} shows the AUC of $P_d$ versus the SNR (in the range of $[0 , 30]$ dB), for each method under both normal and bi-normal noise distributions. 
The first experiment (third column from left) depicts the results of a normal distribution, as seen in Fig.~\ref{fig:pd_vs_snr}. In the second experiment (fourth column from left), we simulate a Bi-normal noise distribution,
\[
    \mathcal{B}_d = \mathbbm{1}_{u<r}\cdot\mathcal{N}(\mu_1,\sigma_1^2) + \mathbbm{1}_{u\geq r}\cdot\mathcal{N}(\mu_2,\sigma_2^2),
\] where $u$ is a random number drawn from the standard uniform distribution, $r$ is a pre-defined threshold set to $r=0.75$, and $(\mu_1,\sigma_1^2) = (0,1)$, $(\mu_2,\sigma_2^2) = (3,0.25)$.
The results demonstrate that REECAP achieves the highest area under the $P_d$ versus SNR curve under both noise models, decreasing from $0.57$ under Gaussian noise to $0.53$ under bi-normal noise. This corresponds to a relative drop of $7.02\%$. Compared to the other a posteriori methods, REECAP is the second least affected after ESTER, which shows no degradation but achieves substantially lower AUC values.
EVT shows a smaller drop of $4.44\%$ while maintaining the closest AUC values to REECAP, whereas AIC, RIVES, SAMP, and SAMP++ show larger drops $15.63\% - 22.22\%$. 
The spectral-gap method remains largely insensitive to these variations, likely due to its reliance solely on spectral thresholding, but exhibits poor detection performance. SDD performs even worse, with AUC values near zero, precluding meaningful analysis.

\begin{table}[t]
  \centering
  \caption{AUC-$P_d$ comparison of the different methods under normal and bi-normal noise distributions.}
  \begin{tabular}{llccc}
    \toprule
    \textbf{Group} & \textbf{Method} & \textbf{Normal}
      & \textbf{Bi-normal} & \textbf{Drop (\%)} \\
    \midrule
    \multirow{2}{*}{\shortstack[l]{Spectral-\\thresholding}}
      & Gap & 0.18 & 0.18 & 0.00 \\
      & SDD & 0.00 & 0.00 & -- \\
    \midrule
    \multirow{2}{*}{ITC-based}
      & EVT & $\underline{0.45}$ & $\underline{0.43}$ & $\underline{4.44}$ \\
      & AIC & 0.32 & 0.27 & 15.63 \\
    \midrule
    \multirow{5}{*}{\shortstack[l]{A posteriori\\methods}}
      & REECAP & $\mathbf{0.57}$ & $\mathbf{0.53}$ & ${7.02}$ \\
      & ESTER  & 0.27 & 0.27 & $\mathbf{0.00}$ \\
      & RIVES  & 0.34 & 0.28 & 17.65 \\
      & SAMP   & 0.27 & 0.21 & 22.22 \\
      & SAMP++ & 0.37 & 0.29 & 21.62 \\
    \bottomrule
  \end{tabular}
  \label{table:pd_auc_noise}
\end{table}

\subsection{Sensitivity Analysis}\label{subsec:results_sensitivity_analysis}

\begin{figure}[t]
    \centering

    \begin{subfigure}[t]{0.48\linewidth}
        \centering
        \includegraphics[width=\linewidth]{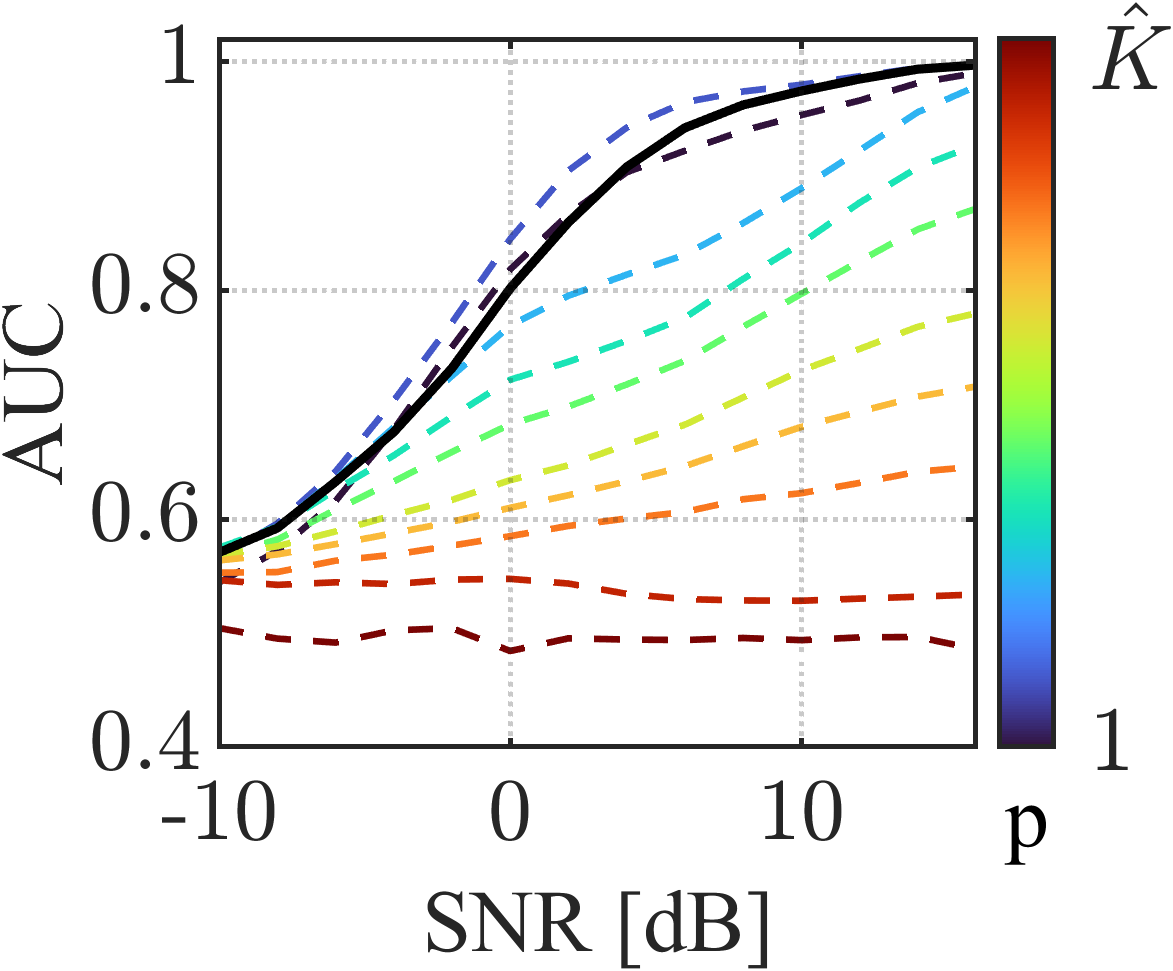}
        \caption{}
        \label{fig:sens_p_vs_SNR}
    \end{subfigure}\hfill
    \begin{subfigure}[t]{0.48\linewidth}
        \centering
        \includegraphics[width=\linewidth]{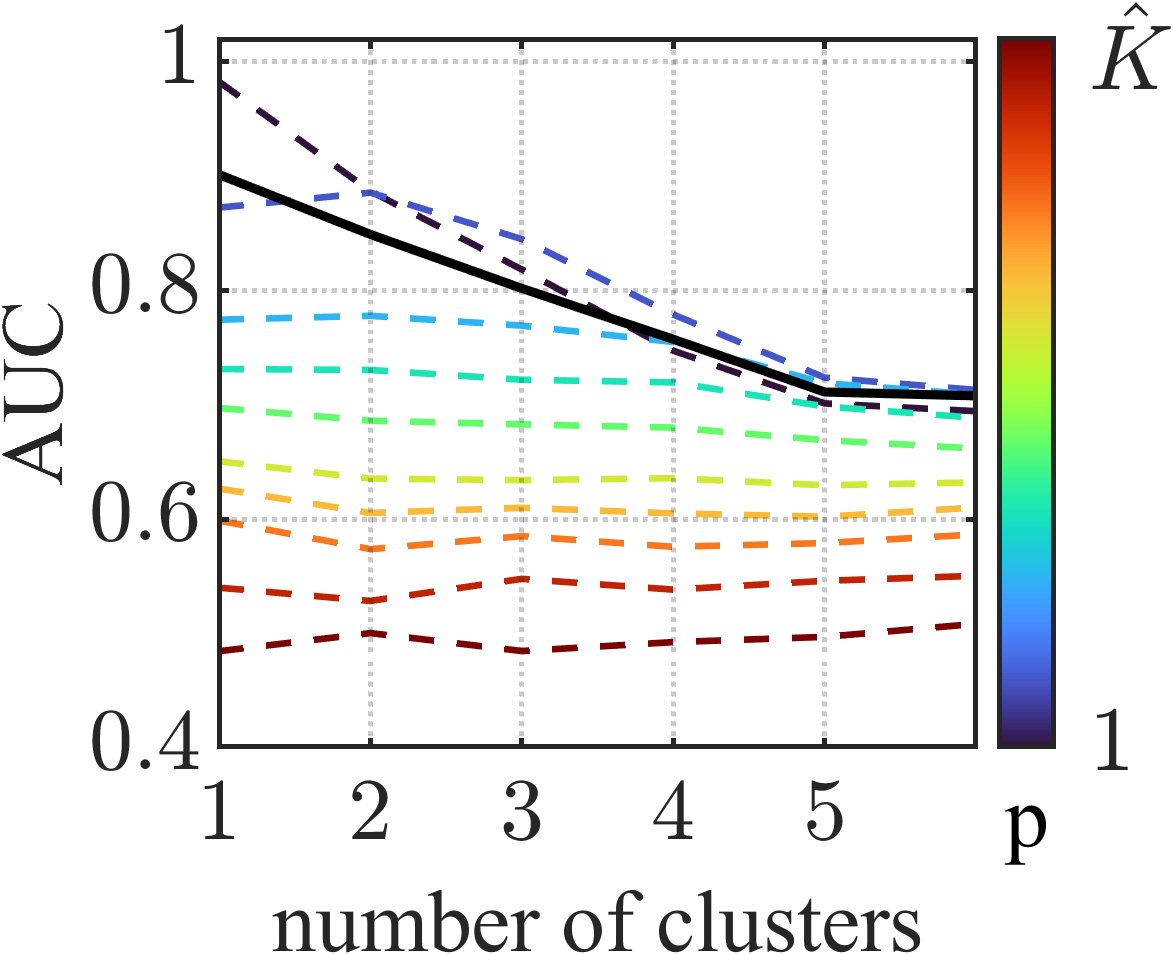}
        \caption{}
        \label{fig:sens_p_vs_nclusters}
    \end{subfigure}

    \par\smallskip

    \begin{subfigure}[t]{0.48\linewidth}
        \centering
        \includegraphics[width=\linewidth]{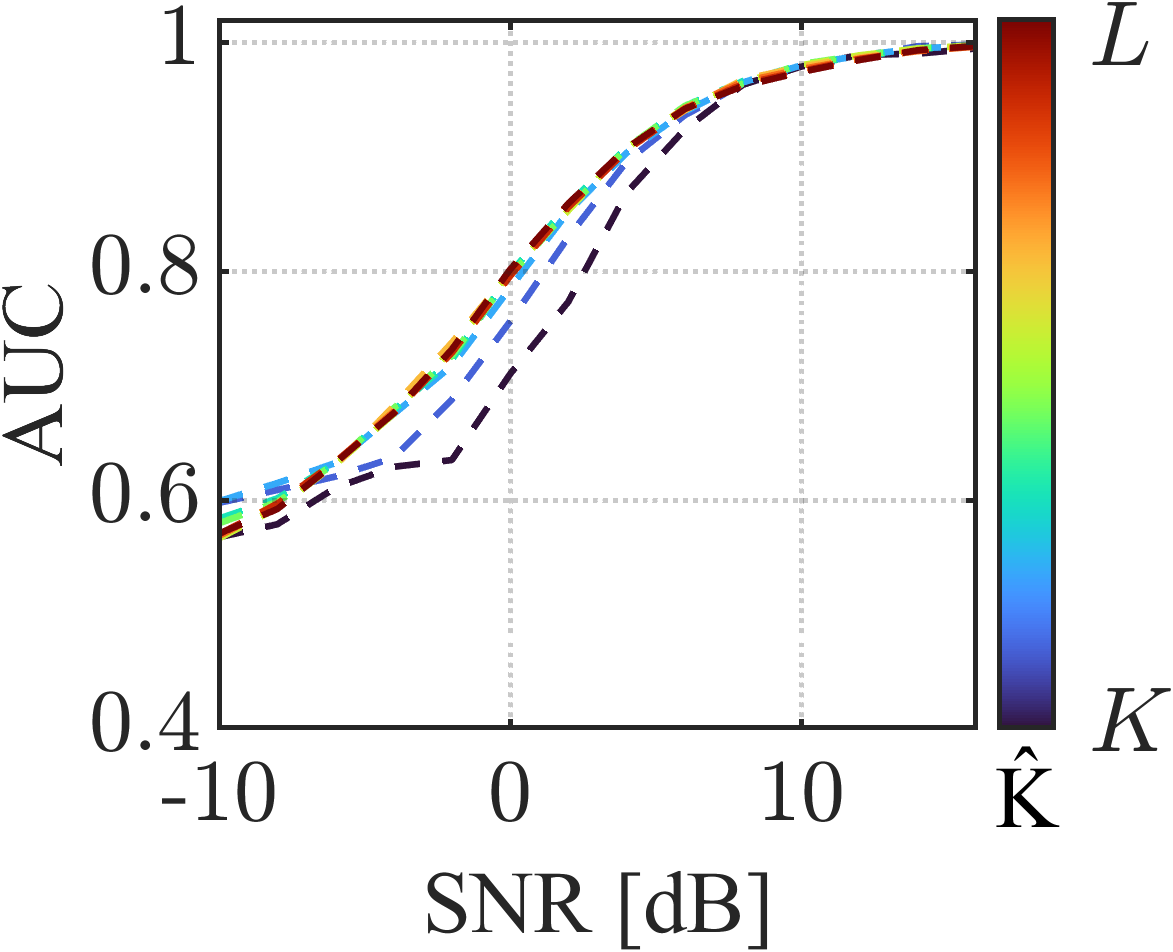}
        \caption{}
        \label{fig:sens_Khat_vs_SNR}
    \end{subfigure}\hfill
    \begin{subfigure}[t]{0.48\linewidth}
        \centering
        \includegraphics[width=\linewidth]{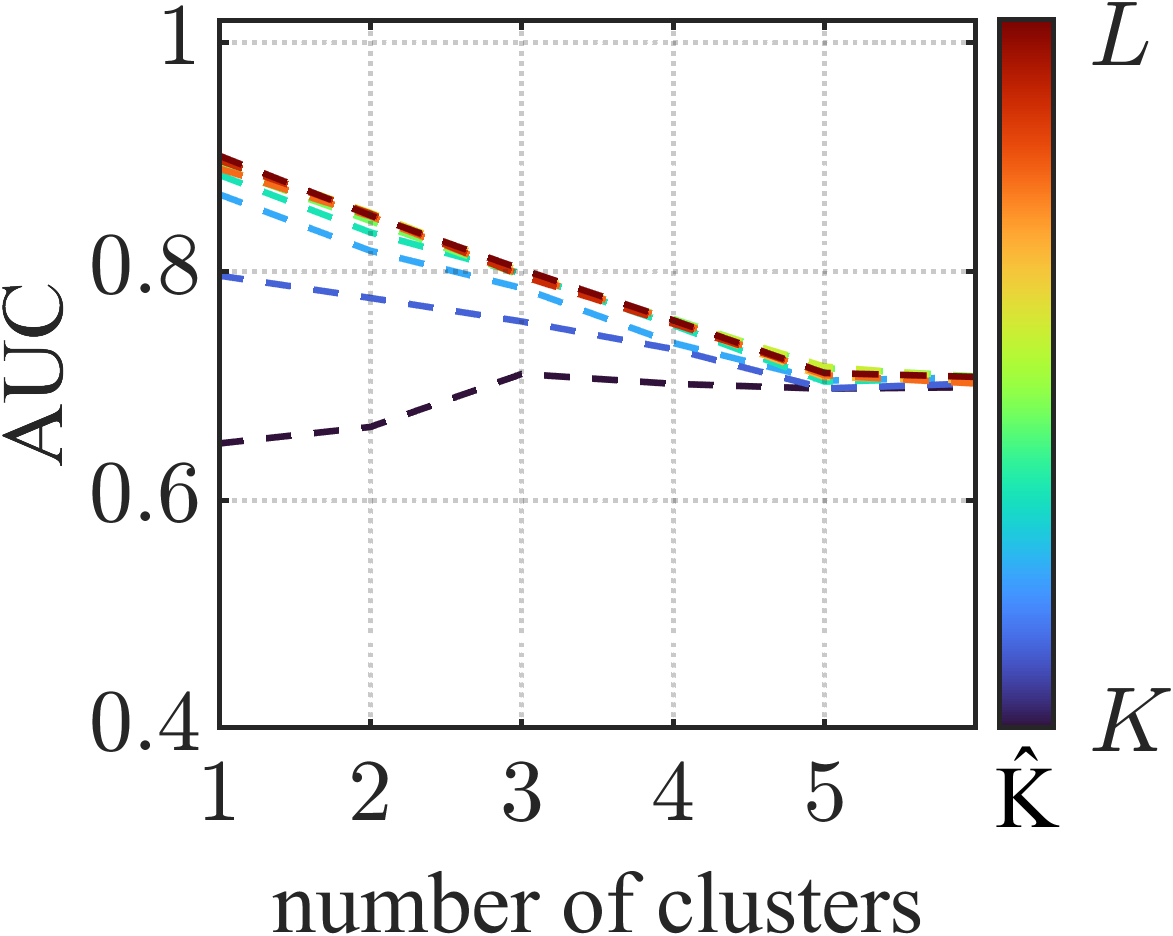}
        \caption{}
        \label{fig:sens_Khat_vs_nclusters}
    \end{subfigure}

    \par\smallskip

    \begin{subfigure}[t]{0.48\linewidth}
        \centering
        \includegraphics[width=\linewidth]{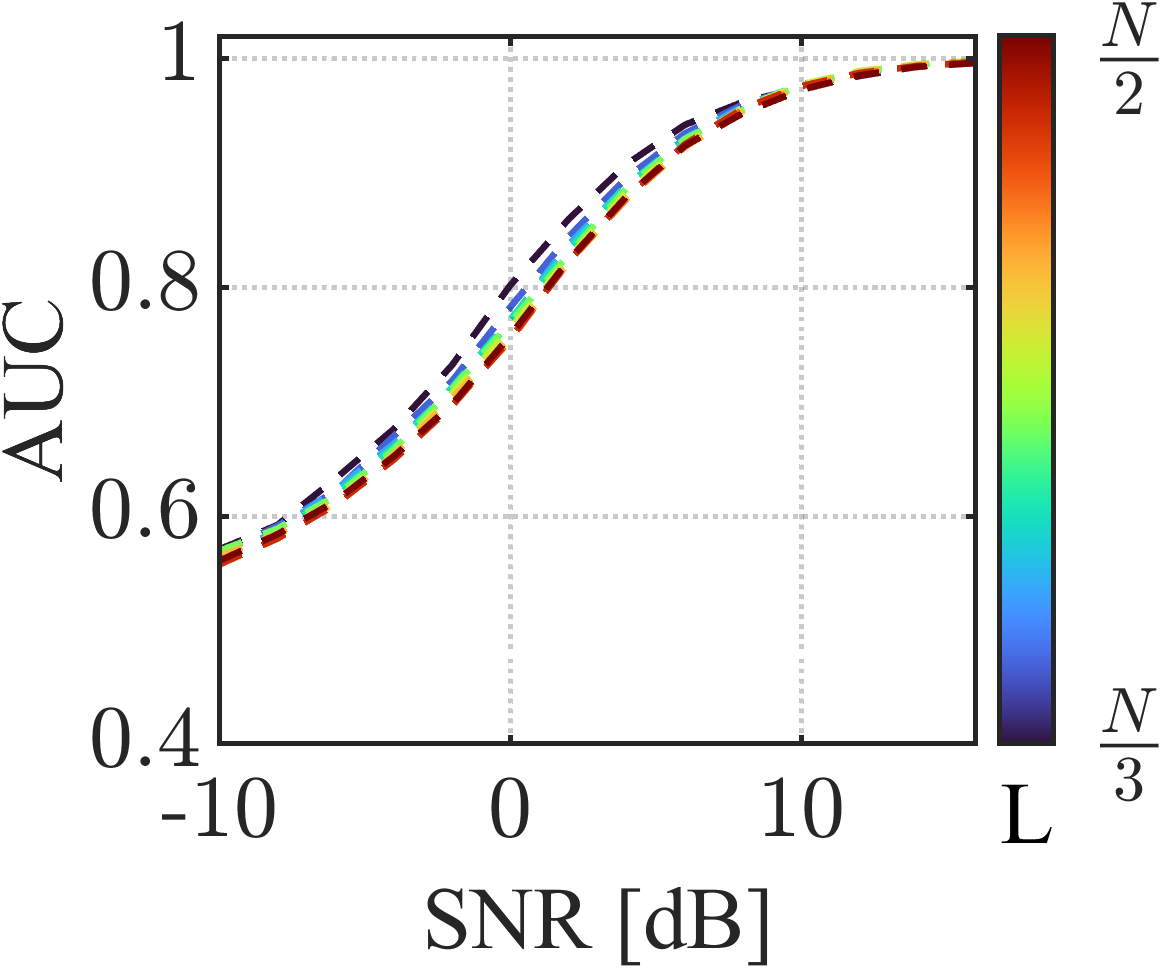}
        \caption{}
        \label{fig:sens_L_vs_SNR}
    \end{subfigure}\hfill
    \begin{subfigure}[t]{0.48\linewidth}
        \centering
        \includegraphics[width=\linewidth]{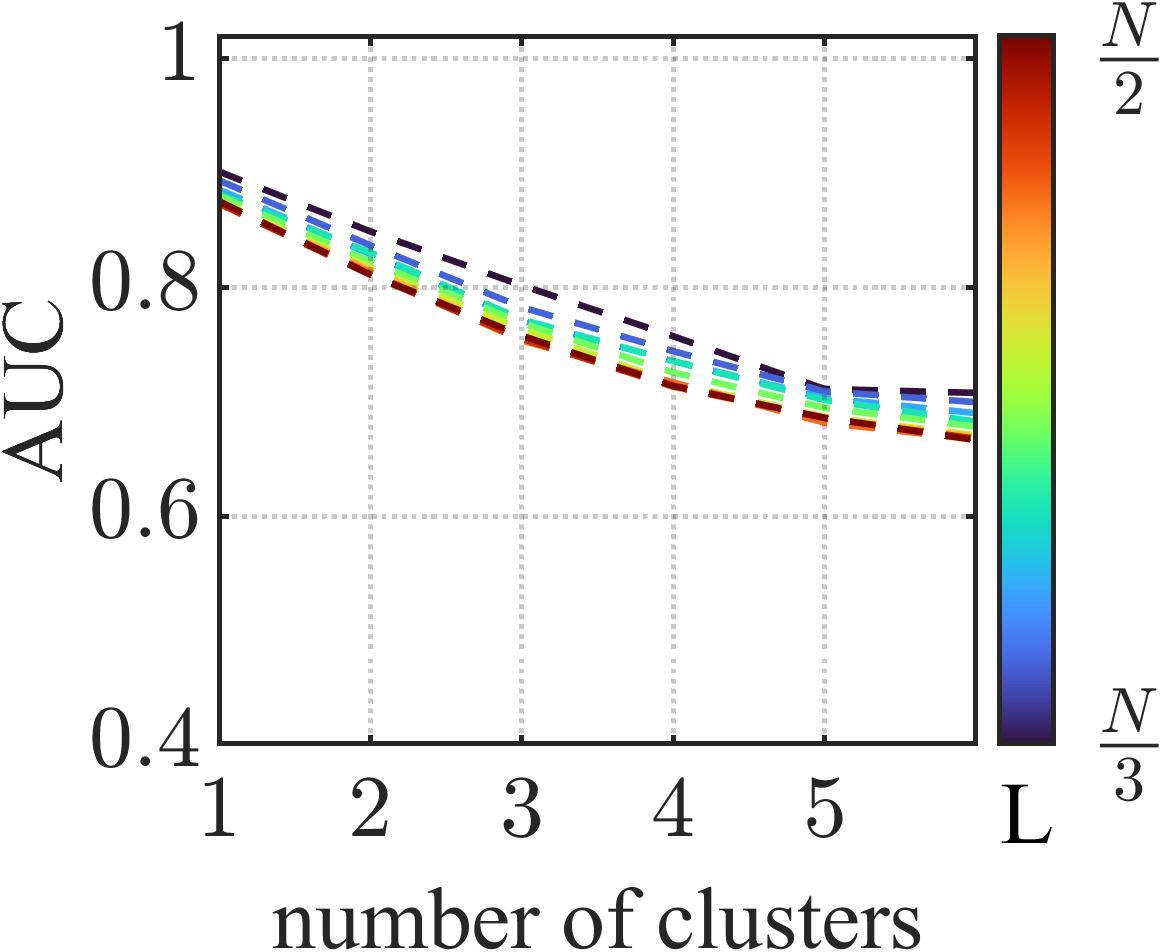}
        \caption{}
        \label{fig:sens_L_vs_nclusters}
    \end{subfigure}

    \caption{AUC-ROC versus SNR (left) and number of clusters
    (right), varying $p$ (top), $\widehat K$ (middle), and $L$
    (bottom). Solid black curves correspond to selecting $p$ using the elbow method.}
    \label{fig:sensitivity_analysis}
\end{figure}

To assess the effect of the hyperparameters in REECAP, we vary the score rank \(p\), the working rank \(\widehat K\), and the pencil parameter \(L\), and evaluate performance using the AUC-ROC. 
Fig.~\ref{fig:sensitivity_analysis} presents the results versus SNR and the number of clusters. The cluster sweep is particularly relevant to $p$, since in our simulations $p$ is chosen to estimate the number of dominant right singular vectors of $\Y_0$, which is naturally related to the number of clusters in our clustered pole configuration. The remaining signal parameters show similar trends, and are presented in the SM.
Figs.~\ref{fig:sens_p_vs_SNR}–\ref{fig:sens_p_vs_nclusters} show that the score rank \(p\) has the strongest impact on the performance of REECAP. The solid black curves correspond to selecting $p$ using the elbow method.
%
%
Figs.~\ref{fig:sens_Khat_vs_SNR}–\ref{fig:sens_Khat_vs_nclusters} show that the choice of $\widehat K$ has a smaller effect, provided that $\widehat K$ is chosen strictly larger than $K$. 
%
Finally, the performance is less sensitive to the choices of \(L\), as the corresponding curves largely overlap. Among the tested values, smaller \(L\) provides a modest advantage.

\subsection{Computational Load}
\begin{figure}[t]
    \centering
    \includegraphics[width=0.93\linewidth]{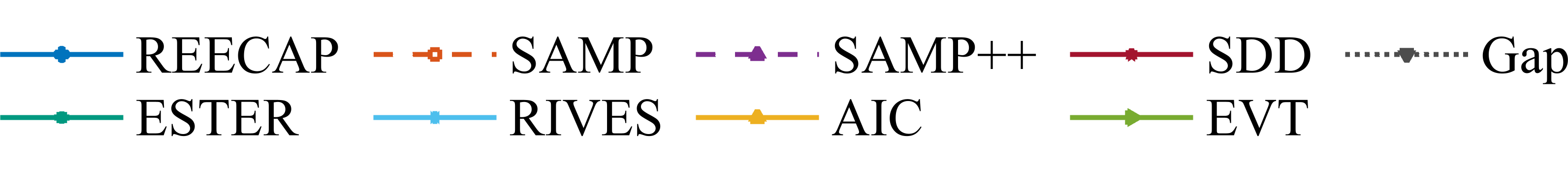}
    \includegraphics[width=0.88\linewidth]{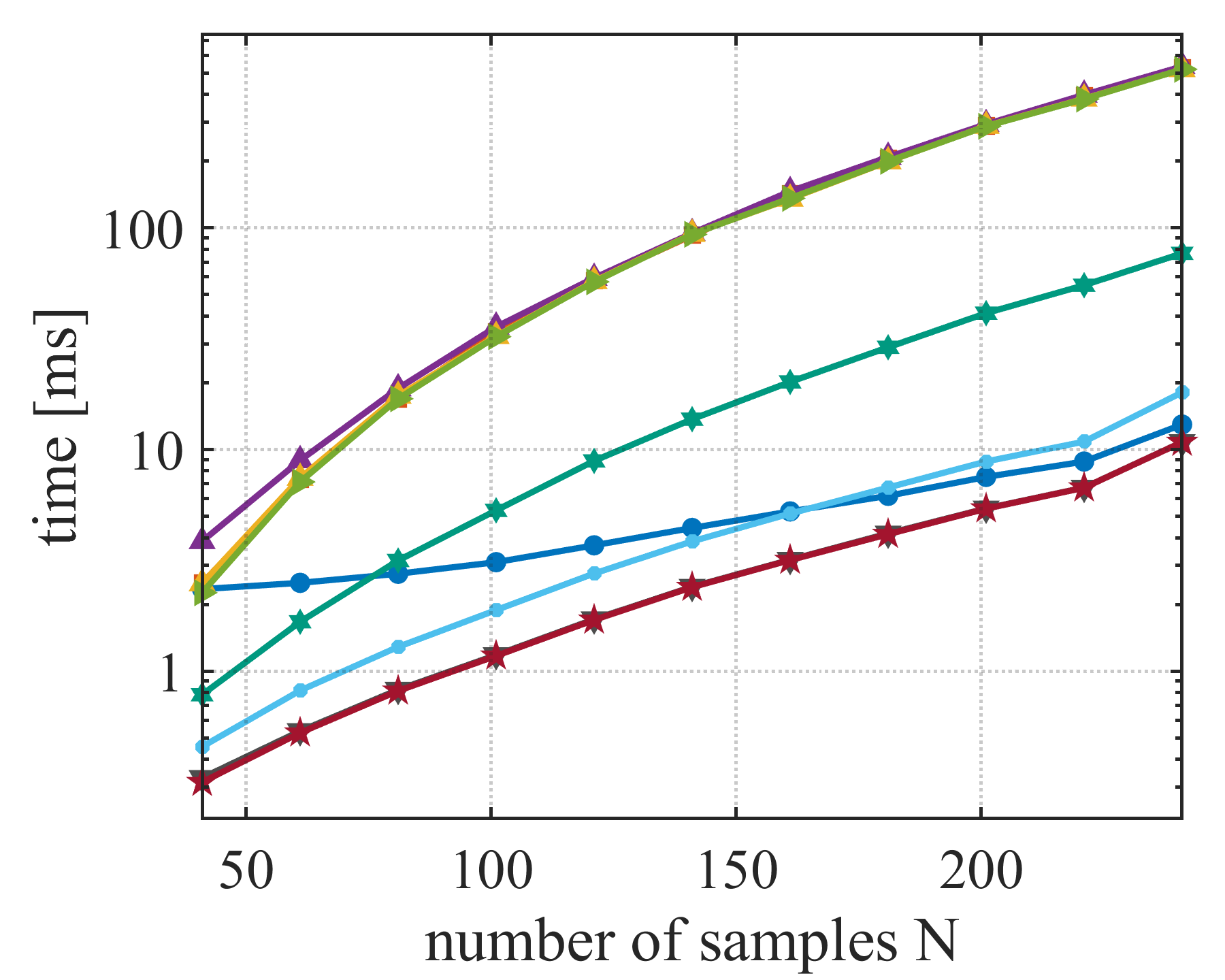}
    \caption{Computational time (log scale) vs number of samples
for all competing methods.}
    \label{fig:comput_time}
\end{figure}

Fig.~\ref{fig:comput_time} compares computation time as a function of the number of samples. It can be seen that REECAP is substantially more efficient than SAMP, SAMP++, AIC, and EVT, but generally slower than the spectral-thresholding methods Gap and SDD. It is faster than ESTER for all but the two smallest tested sample sizes, and relative to RIVES, it is slower for short records, and comparable for longer records. 
Specifically, the proposed method requires, on average, approximately $21.5$ times faster than the ITC-based and SAMP methods, and approximately $3$ times slower than the spectral thresholding methods.
This behavior is expected: ITC-based methods require evaluating the likelihood function over all candidate model orders, whereas spectral thresholding requires only \(\mathcal O(L)\) additional operations. This is lower than the \(\mathcal O((p + RT)\widehat K)\) additional operations of REECAP, as detailed in Section~\ref{susec:comput_complex}. 

\section{Conclusion}
This paper introduced REECAP, a new MP-based algorithm for model order detection and pole estimation of sums of damped complex exponentials in noise. REECAP adopts an a posteriori approach, evaluating candidate poles using a score that measures the concentration of their associated eigenvector energy in the dominant right singular subspace of the data Hankel matrix. Our analysis establishes a deterministic sufficient condition connecting this score to pole-estimation accuracy: under a suitable rank condition, exceeding a certain threshold guarantees proximity to at least one signal pole. This provides a theoretical foundation for the proposed score.

Numerical experiments demonstrated improved component classification, model order detection, and frequency estimation performance across a broad range of signal configurations. REECAP requires little computation beyond the conventional MP method and achieved lower runtimes than the evaluated structure-aware MP and ITC-based methods. Sensitivity analysis identified the score rank as the most influential hyperparameter, with the singular-value elbow providing a simple and effective rule for its selection in the considered scenarios.

Future work will focus on constructing thresholds computable from the observations that retain estimation guarantees, and on establishing conditions for distinct-pole recovery and correct model order detection.

\section*{Acknowledgment}
The authors thank Yoav Harris for valuable discussions and helpful comments on the manuscript.

The authors used OpenAI Codex and Anthropic Claude Code to assist in writing and debugging the simulation code underlying the numerical results. The authors take full responsibility for the implementation and reported results.

\printbibliography
\end{refsection}
\newpage
\makeatletter
\setlength{\@fptop}{0pt}
\setlength{\@dblfptop}{0pt}
\makeatother
\twocolumn[
\begin{center}

    {\huge Model Order Detection with Parameter Estimation Guarantees for the Matrix Pencil Method\\  \vspace{0.2cm}
    \textbf{Supplementary Materials}}
    
    \vspace{0.5cm}
    
        {Yehonatan-Itay Segman\textsuperscript{1,$\dagger$},~
        \IEEEmembership{Student Member,~IEEE},
        David Sadka\textsuperscript{2,$\dagger$},
        and Ronen Talmon\textsuperscript{1},~\IEEEmembership{Senior Member,~IEEE}
        }
\end{center}
]

\setcounter{page}{1}
{
\appendices
\begin{NoHyper}
\begin{refsection}

\appendices
\section{Proof of Proposition~\ref{prop:Score_implications}}
\label{app:Score_implications_proof}
\renewcommand\thefootnote{}
\footnote{\textsuperscript{$\dagger$}Contributed equally.}%
\footnote{\textsuperscript{1}Viterbi Faculty of Electrical and Computer Engineering, Technion -- Israel Institute of Technology, Haifa 32000, Israel; yehonatans@campus.technion.ac.il, ronen@ee.technion.ac.il.}%
\footnote{\textsuperscript{2}Department of Electrical and Computer Engineering, Yale University, New Haven, CT, USA; david.sadka@yale.edu.}%
\renewcommand\thefootnote{\arabic{footnote}}
\addtocounter{footnote}{-1}



We start by showing that $\mu_{\bDelta}>0 $ is equivalent to $\operatorname{Rank}(\U_{\widehat K}^{\mathrm H}\Z_L)=K$. Since $\Z_R$ has full row rank, $\sigma_K(\Z_R)>0$. Moreover, $\B\boldsymbol{\Delta}$ is
invertible because $b_k\neq0$ and $\Delta_k>0$ for every $k$.
Consequently,
\[
\begin{aligned}
\mu_{\Delta}>0
&\iff
\sigma_K\!\left(
\U_{\widehat K}^{\mathrm H}\Z_L\B\boldsymbol{\Delta}
\right)>0\\
&\iff
\operatorname{Rank}\!\left(
\U_{\widehat K}^{\mathrm H}\Z_L\B\boldsymbol{\Delta}
\right)=K\\
&\iff
\operatorname{Rank}\!\left(
\U_{\widehat K}^{\mathrm H}\Z_L
\right)=K,
\end{aligned}
\]
where the middle equivalence follows since the matrix $\U_{\widehat K}^{\mathrm H}\Z_L\B\boldsymbol{\Delta}$ has $K$ columns. The last equivalence follows because multiplying by \(\B\boldsymbol{\Delta}\) only rescales the columns.

We now turn to prove Proposition~\ref{prop:Score_implications}. We prove the contrapositive, and suppose \(\lambda_i\notin D\). Starting from the projected equation in~\eqref{eq:projected_MP_equation}
\[
    \U_{\widehat K}^{\mathrm{H}}(\Y_1 - \lambda_i \Y_0) \mathbf d_i = 0.
\]
Since $\Y_0 = \X_0 + \W_0$, and $\Y_1 = \X_1 + \W_1$, we get
\begin{equation}\label{eq:some_eq}
   \|\U_{\widehat K}^{\mathrm{H}}(\X_1-\lambda_i \X_0) \mathbf d_i \| =  \|\U_{\widehat K}^{\mathrm{H}}(\W_1-\lambda_i \W_0)\mathbf d_i\|.
\end{equation}

Using the noiseless Vandermonde factorization 
\[
    \X_1- \lambda_i \X_0 = \Z_L \B (\Z-\lambda_i \I)\Z_R, 
\]
the noiseless projected equation can be recast as
\begin{align*}
    &\U_{\widehat K}^{\mathrm{H}}(\X_1-\lambda_i \X_0) \mathbf d_i = \\
    &\U_{\widehat K}^{\mathrm{H}} \Z_L \B \boldsymbol{\Delta}\left[\boldsymbol{\Delta}^{-1}(\Z-\lambda_i \I)\right]\Z_R \mathbf d_i.
\end{align*}

Since the matrix $\boldsymbol{\Delta}^{-1}(\Z-\lambda_i \I)$ is diagonal, its smallest singular values equals $d_{\cD}(\lambda_i)$ defined in ~\eqref{eq:lambda_dist_to_D}. Therefore, since $\|\mathbf d_i\|=1$, $\text{rank}(\Z_L \B \boldsymbol{\Delta}) = K$, and $\widehat K\geq K$,
\begin{align*}
    \|\U_{\widehat K}^{\mathrm{H}}(\X_1-\lambda_i \X_0) \mathbf d_i \| \geq \sigma_K(\U_{\widehat K}^{\mathrm{H}} \Z_L \B \boldsymbol{\Delta}) d_{\cD}(\lambda_i) \| \Z_R \mathbf d_i\|.
\end{align*}

Denote the orthogonal projector on the noiseless right signal subspace by $\mathbf P_x := \Z_R^\dagger \Z_R$. Since $\Z_R \Z_R^{\dagger} = \I$, 
\begin{align*}
\|\Z_R\mathbf d_i\| &= \|\Z_R \Z_R^{\dagger}\Z_R \mathbf d_i\| \\&=\| \Z_R \mathbf P_x \mathbf d_i\|\\ &\geq \sigma_{K}(\Z_R)\|\mathbf P_x \mathbf d_i\|,
\end{align*}
where the rightmost inequality holds since $\mathbf P_x \mathbf d_i \in \operatorname{Range}(\Z_R^{\mathrm{H}})$.
Combining the above inequalities yields
\[
    \mu_{\boldsymbol{\Delta}} d_{\cD}(\lambda_i) \|\mathbf P_x \mathbf d_i\| \leq \|\U_{\widehat K}^{\mathrm{H}}(\W_1-\lambda_i \W_0)\mathbf d_i\|,
\]
since $\mu_{\boldsymbol{\Delta}} = \sigma_K(\U_{\widehat K}^{\mathrm{H}} \Z_L \B \boldsymbol{\Delta})\sigma_K(\Z_R)$. Equivalently
\begin{equation}
    \|\mathbf P_x \mathbf d_i\| \leq \frac{ \|\U_{\widehat K}^{\mathrm{H}}(\W_1-\lambda_i \W_0)\mathbf d_i\|}{\mu_{\boldsymbol{\Delta}} d_{\cD}(\lambda_i)},
\end{equation}
since $\mu_{\boldsymbol{\Delta}}>0$ by assumption, and $ d_{\cD}(\lambda_i) >0$ since we assumed $\lambda_i \notin \mathcal D$.

We now relate $\|\mathbf P_x \mathbf d_i\|$ to $s_p(i)$. 
%
%
Decomposing $\mathbf d_i=\mathbf P_x\mathbf d_i+(\I-\mathbf P_x)\mathbf d_i$, we obtain
\begin{align*}
    \sqrt{s_p(i)} &= \|\V_p\V_p^H\mathbf d_i\|\\
    &\leq
      \|\V_p^{\mathrm{H}} \mathbf P_x\mathbf d_i\|
      +\|\V_p^{\mathrm{H}}(I-\mathbf P_x)\mathbf d_i\|\\
    &\leq
      \|\mathbf P_x\mathbf d_i\|+\|\V_p^{\mathrm{H}}(I-\mathbf P_x)\mathbf d_i\|,
\end{align*}
where we used the fact that the columns of $\V$, and hence of $\V_p$, are orthonormal.
Hence
\[
    \sqrt{s_p(i)} \leq \|\V_p^{\mathrm{H}}(I-\mathbf P_x)\mathbf d_i\| +\frac{ \|\U_{\widehat K}^{\mathrm{H}}(\W_1-\lambda_i \W_0)\mathbf d_i\|}{\mu_{\boldsymbol{\Delta}} d_{\cD}(\lambda_i)}.
\]
Denoting
\[
\Theta_{i,p} = \|\V_p^{\mathrm{H}}(I-\mathbf P_x)\mathbf d_i\| +\frac{ \|\U_{\widehat K}^{\mathrm{H}}(\W_1-\lambda_i \W_0)\mathbf d_i\|}{\mu_{\boldsymbol{\Delta}}},
\]
it holds that 
\[
    d_{\cD}(\lambda_i) > 1 \Rightarrow
    \sqrt{s_p(i)} \leq \Theta_{i,p},
\]
and hence, taking the contrapositive gives
\begin{equation}
\begin{aligned}
    \sqrt{s_p(i)} > \Theta_{i,p} \ \Rightarrow \ d_{\cD}(\lambda_i) & \le 1
\end{aligned}
\end{equation}
which means that  $\lambda_i \in D$.

\section{Comparison of Heuristics for Selecting $p$} \label{app:heuristics_for_selecting_p}

In our simulations, we select \(p\) using an elbow method~\cite{cattell1966scree} applied to the singular values of \(\Y_0\) to estimate the dimension of its dominant right singular subspace. The elbow is identified as the point farthest from the line connecting the first and last points of the ordered singular-value curve, with the preceding values counted as dominant. 
We compare this choice with four practical alternatives: (i) the acceleration-factor method~\cite{raiche2013non}, which identifies the elbow as the point of maximum discrete second derivative of the singular-value sequence; (ii) counting significant peaks in the Fourier spectrum of the observed signal \(y(n)\), defined as peaks whose heights are at least half the maximum peak height; (iii) the SDD criterion~\cite{sarkar1995using}; and (iv) the spectral-gap criterion~\cite{yin2011model}. We also include \(p\) equal to the true number of pole clusters as a reference.

Fig.~\ref{fig:p_heuristics_comp} is similar to Fig.~\ref{fig:AUC_baseline_compare}, but presents the reference choice of \(p\) equal to the number of clusters, with the practical heuristics for selecting \(p\). As expected, using the true number of clusters achieves the best performance. The elbow-based methods yield comparable results, with the acceleration-based method performing slightly better than the standard elbow method. These results show that the dominant singular values of \(\Y_0\) provide an effective practical surrogate for selecting \(p\) when the number of clusters is unknown.
However, further research is needed to investigate optimal strategies for selecting the score rank $p$ to optimize the tradeoff between reliable component classification and estimation accuracy across different signal configurations.

\begin{figure}[t]
    \centering
    \includegraphics[width=\linewidth]{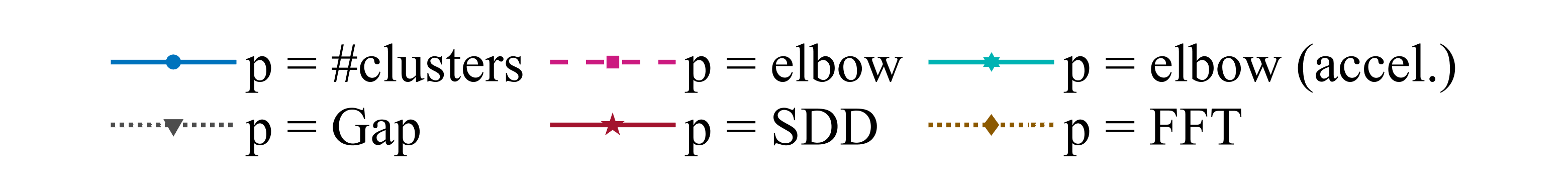}

    \begin{subfigure}[t]{0.48\linewidth}
        \centering
        \includegraphics[width=\linewidth]{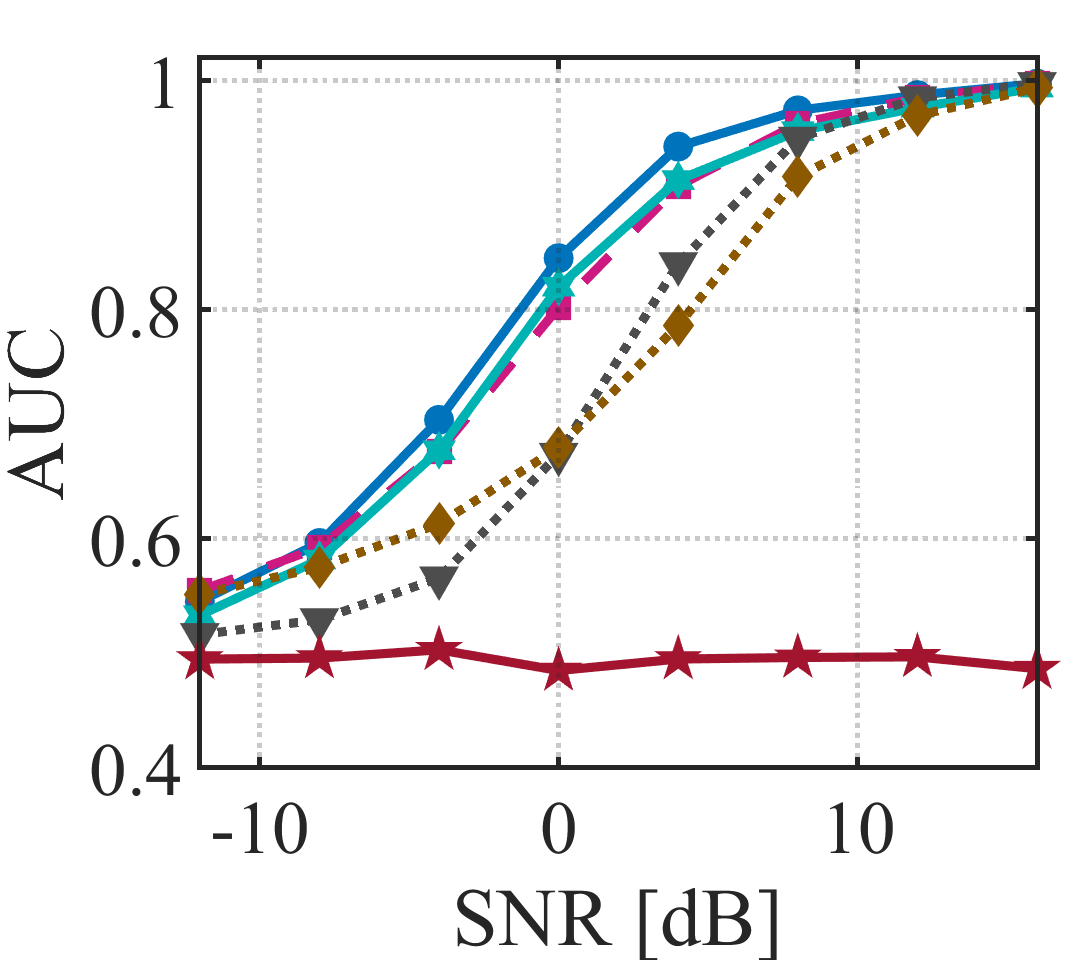}
        \caption{}
        \label{}
    \end{subfigure}\hfill
    \begin{subfigure}[t]{0.48\linewidth}
        \centering
        \includegraphics[width=\linewidth]{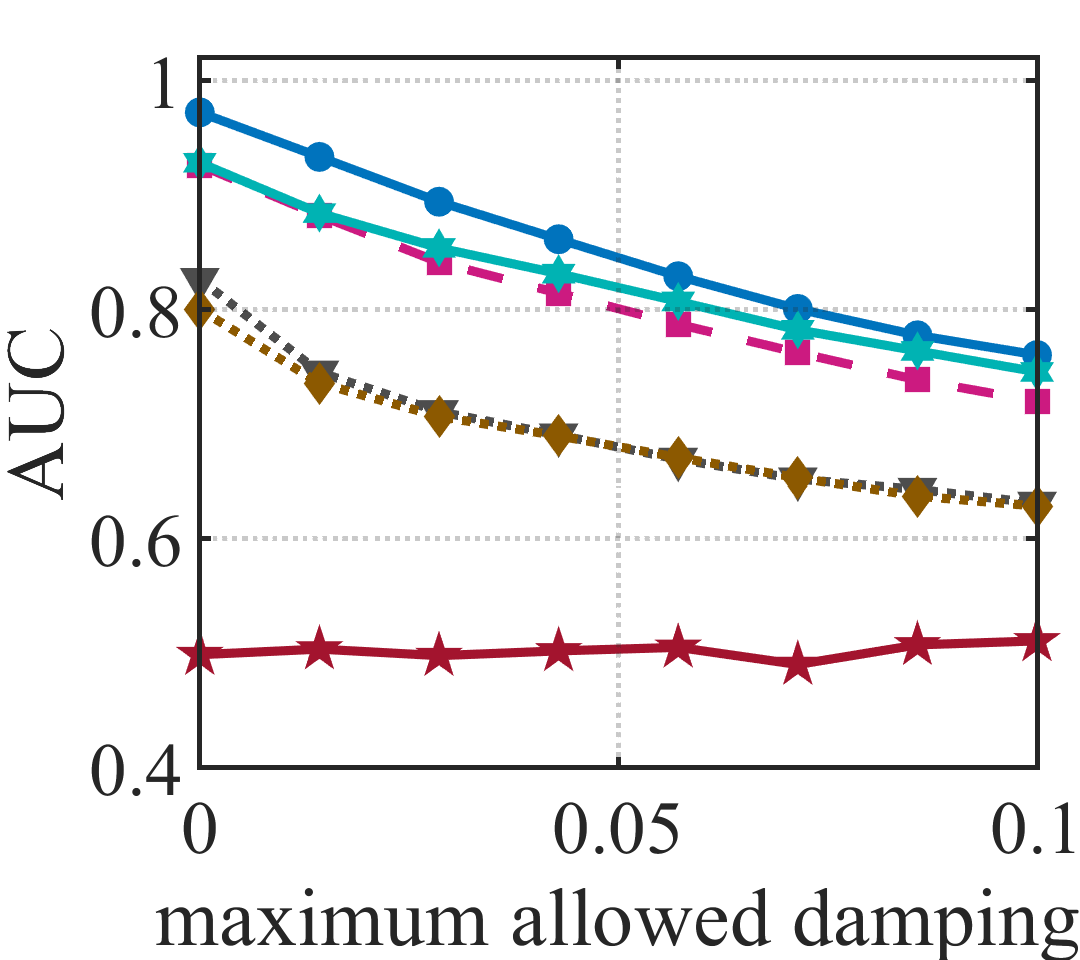}
        \caption{}
        \label{}
    \end{subfigure}


    \begin{subfigure}[t]{0.48\linewidth}
        \centering
        \includegraphics[width=\linewidth]{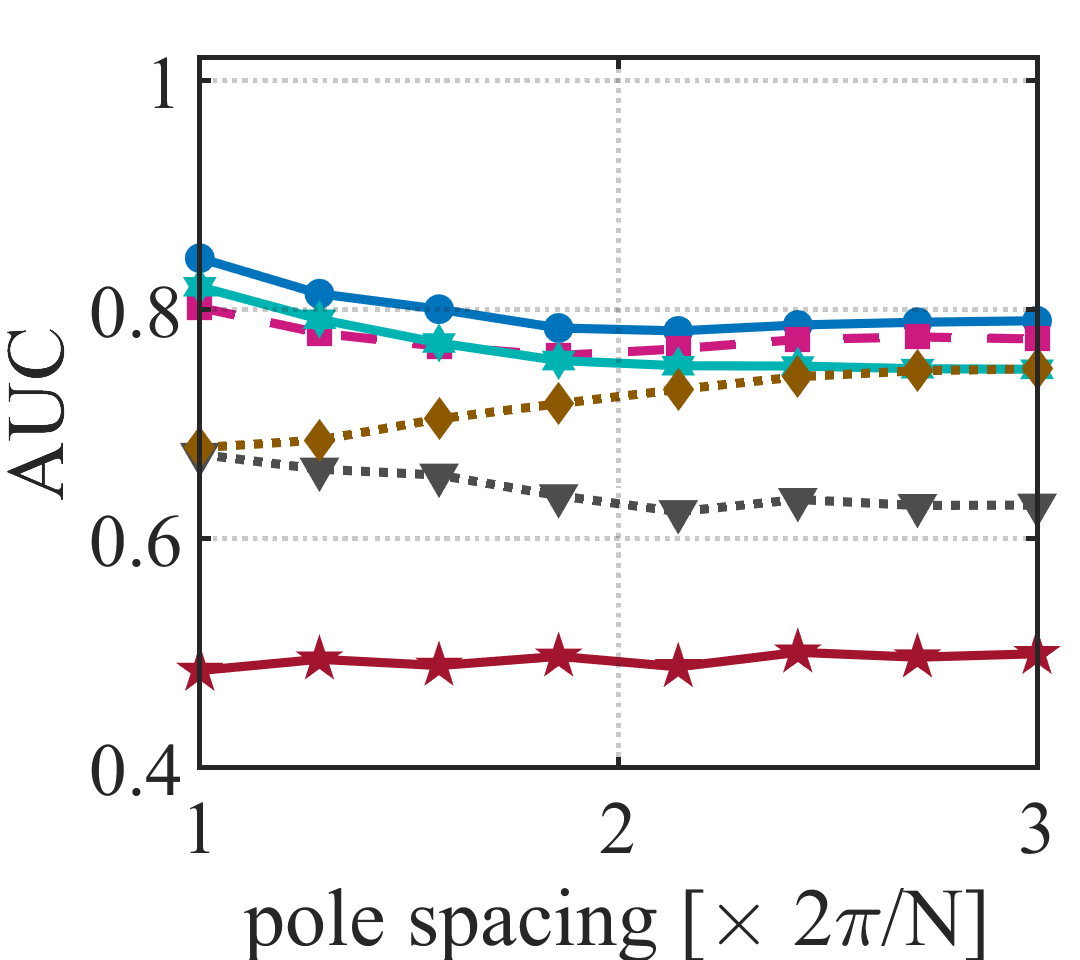}
        \caption{}
        \label{}
        \label{}
    \end{subfigure}\hfill
    \begin{subfigure}[t]{0.48\linewidth}
        \centering
        \includegraphics[width=\linewidth]{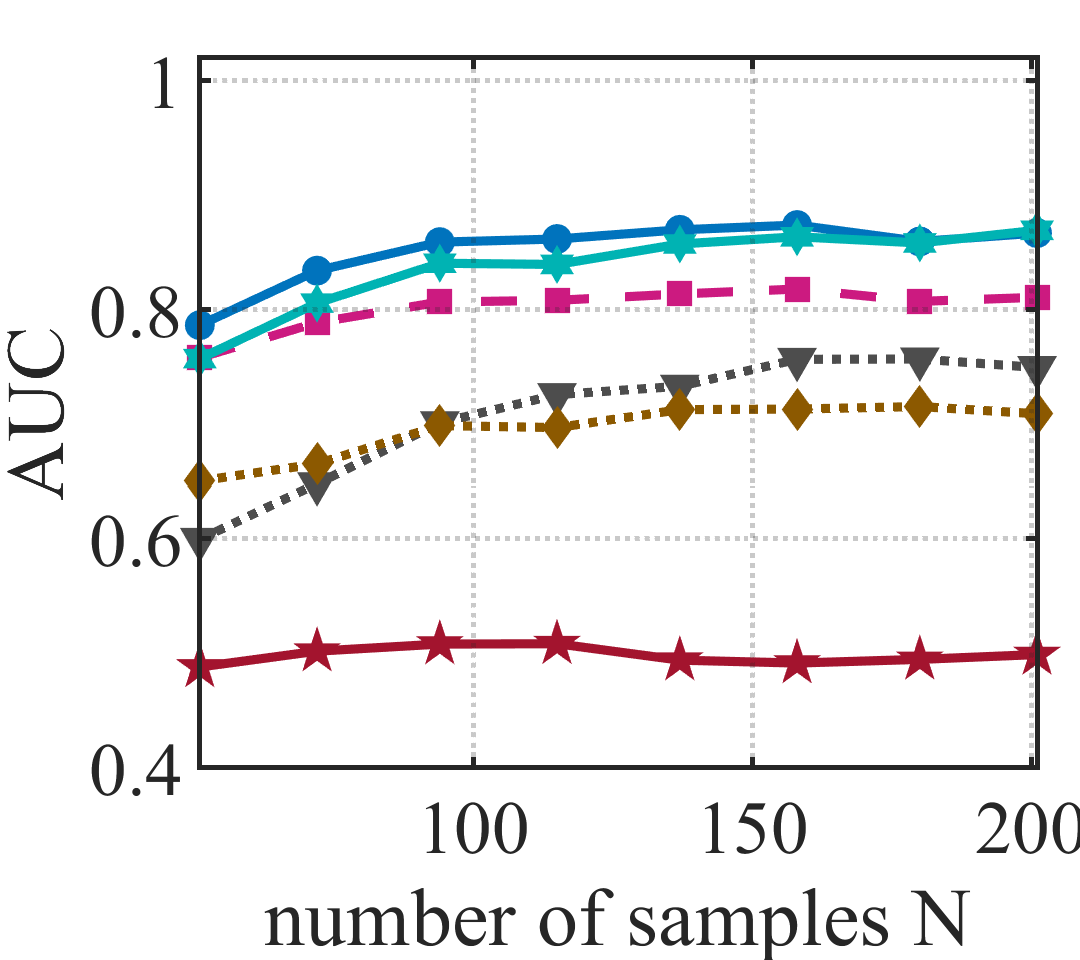}
        \caption{}
        \label{}
    \end{subfigure}


    \begin{subfigure}[t]{0.48\linewidth}
        \centering
        \includegraphics[width=\linewidth]{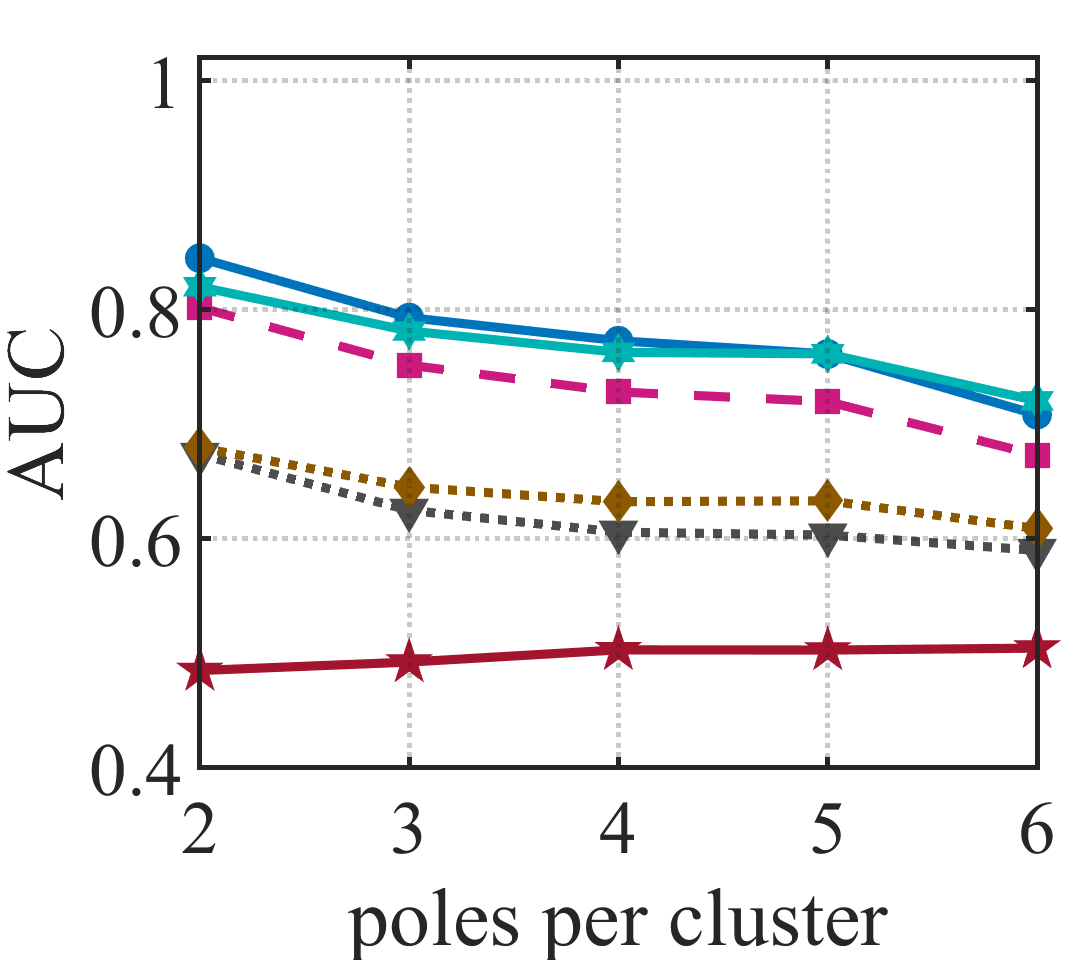}
        \caption{}
        \label{}
    \end{subfigure}\hfill
    \begin{subfigure}[t]{0.48\linewidth}
        \centering
        \includegraphics[width=\linewidth]{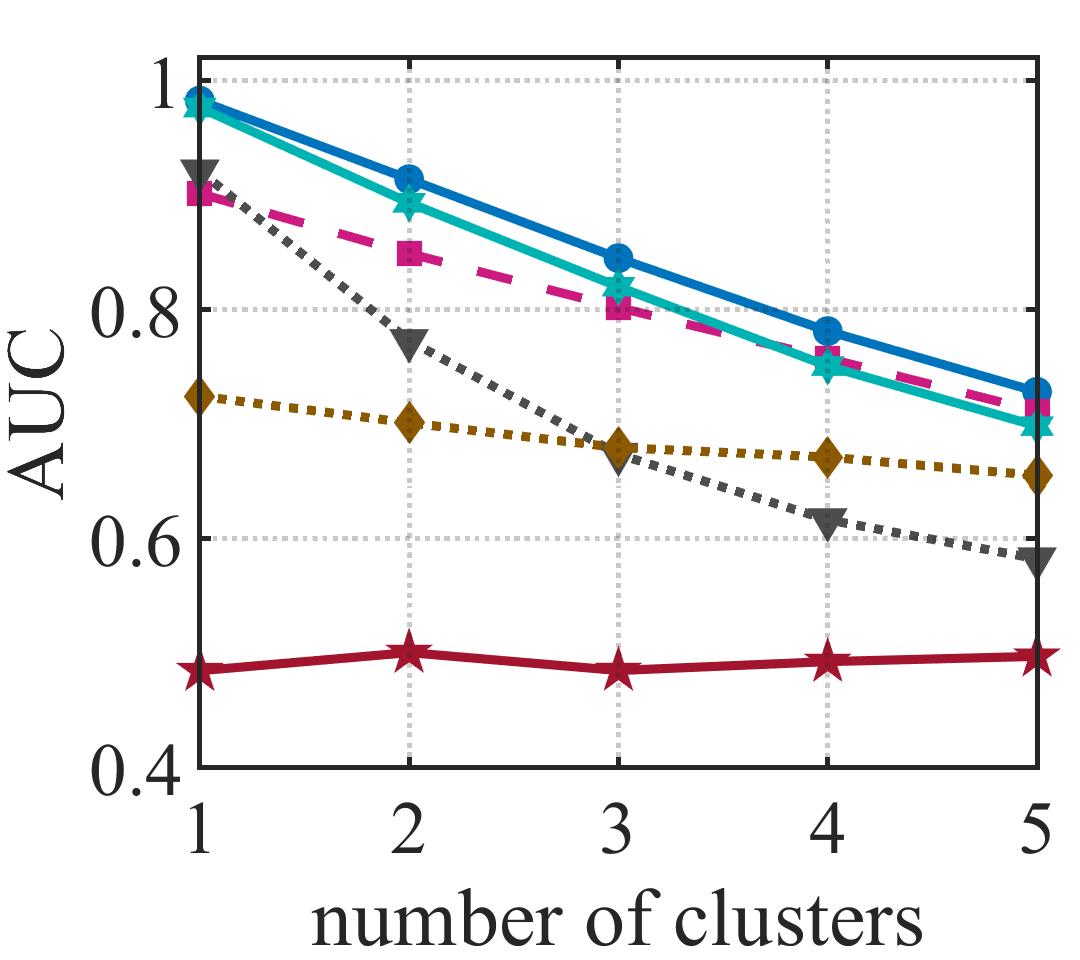}
        \caption{}
        \label{}
    \end{subfigure}
    \caption{AUC-ROC comparison of the proposed REECAP algorithm for different $p$ estimation methods across the simulation parameters.}
    \label{fig:p_heuristics_comp}
\end{figure}

\section{Additional Numerical Results}\label{app:additional_results}
In this appendix, we provide the complementary results for Section~\ref{subsec:results_sensitivity_analysis}. Figs.\ref{fig:sensitivity_analysis_complementary} and \ref{fig:sensitivity_analysis_complementary2}, complement Fig.~\ref{fig:sensitivity_analysis}, and assess the effect of the hyperparameters in REECAP using the AUC-ROC versus the maximum allowed damping per pole, pole-spacing, number of samples, and number of poles per cluster. 
Similarly to Fig.~\ref{fig:sensitivity_analysis}, the score rank \(p\) has the strongest impact on REECAP performance. The solid black curves correspond to selecting \(p\) using the elbow method, whose performance is suboptimal relative to the best tested choices, indicating that an alternative selection of \(p\) could further improve REECAP performance. The choice of \(\widehat K\) has a moderate impact, provided that \(\widehat K\) is strictly larger than \(K\), particularly for the number of samples. Finally, \(L\) has little effect on performance, with a modest advantage for smaller values.

\begin{figure}[t]
    \centering

    \begin{subfigure}[t]{0.49\linewidth}
        \centering
        \includegraphics[width=\linewidth]{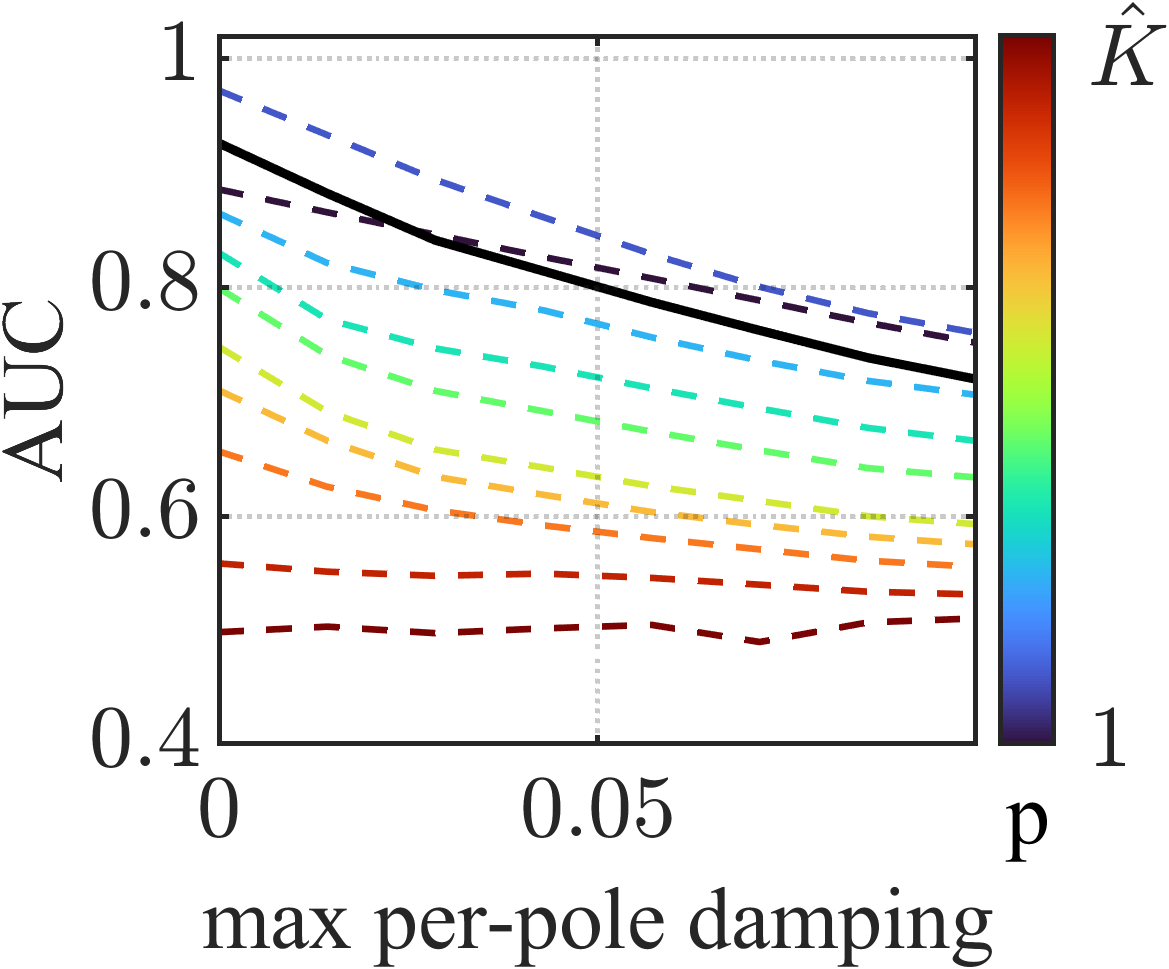}
        \caption{}
        \label{fig:sens_p_vs_damping}
    \end{subfigure}\hfill
    \begin{subfigure}[t]{0.49\linewidth}
        \centering
        \includegraphics[width=\linewidth]{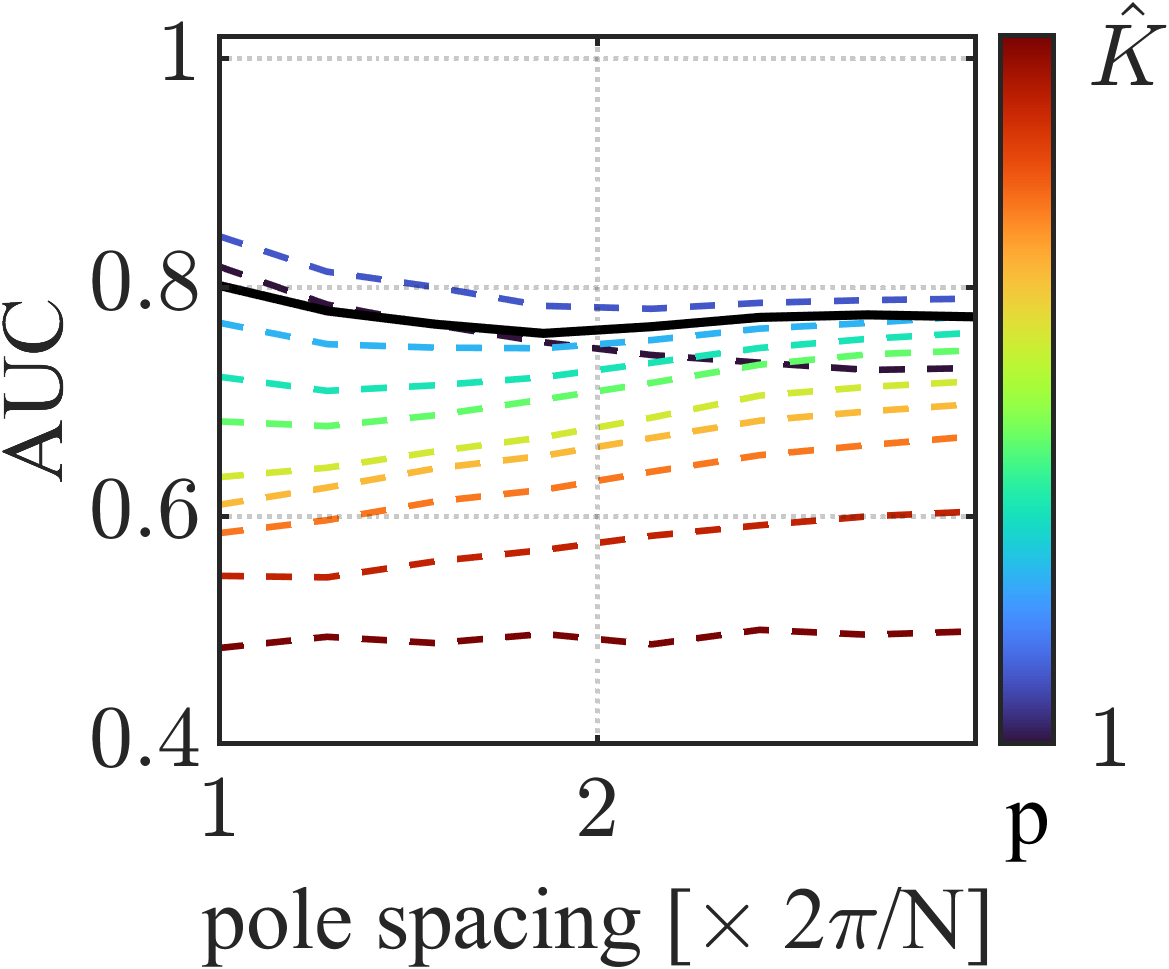}
        \caption{}
        \label{fig:sens_p_vs_minSepR}
    \end{subfigure}

    \par\smallskip

    \begin{subfigure}[t]{0.49\linewidth}
        \centering
        \includegraphics[width=\linewidth]{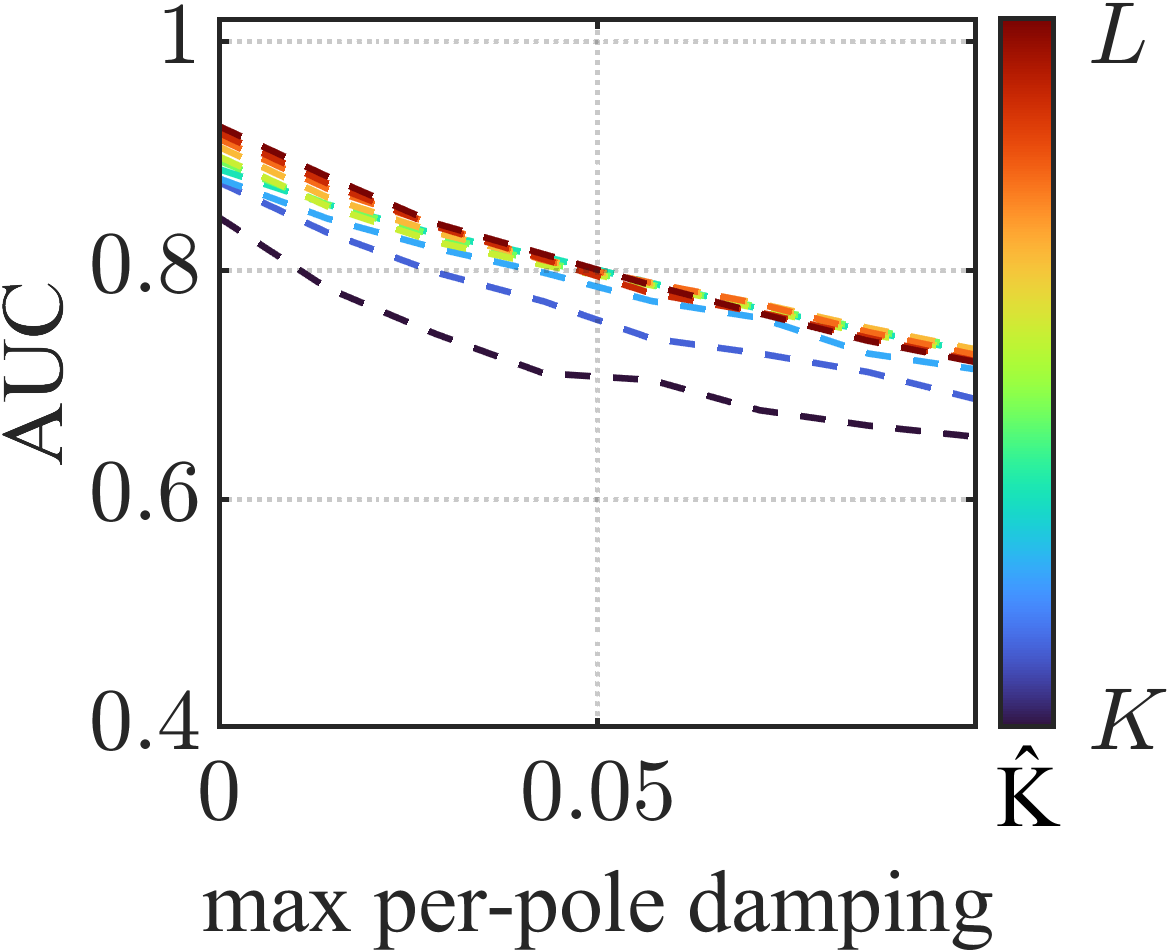}
        \caption{}
        \label{fig:sens_Khat_vs_damping}
    \end{subfigure}\hfill
    \begin{subfigure}[t]{0.49\linewidth}
        \centering
        \includegraphics[width=\linewidth]{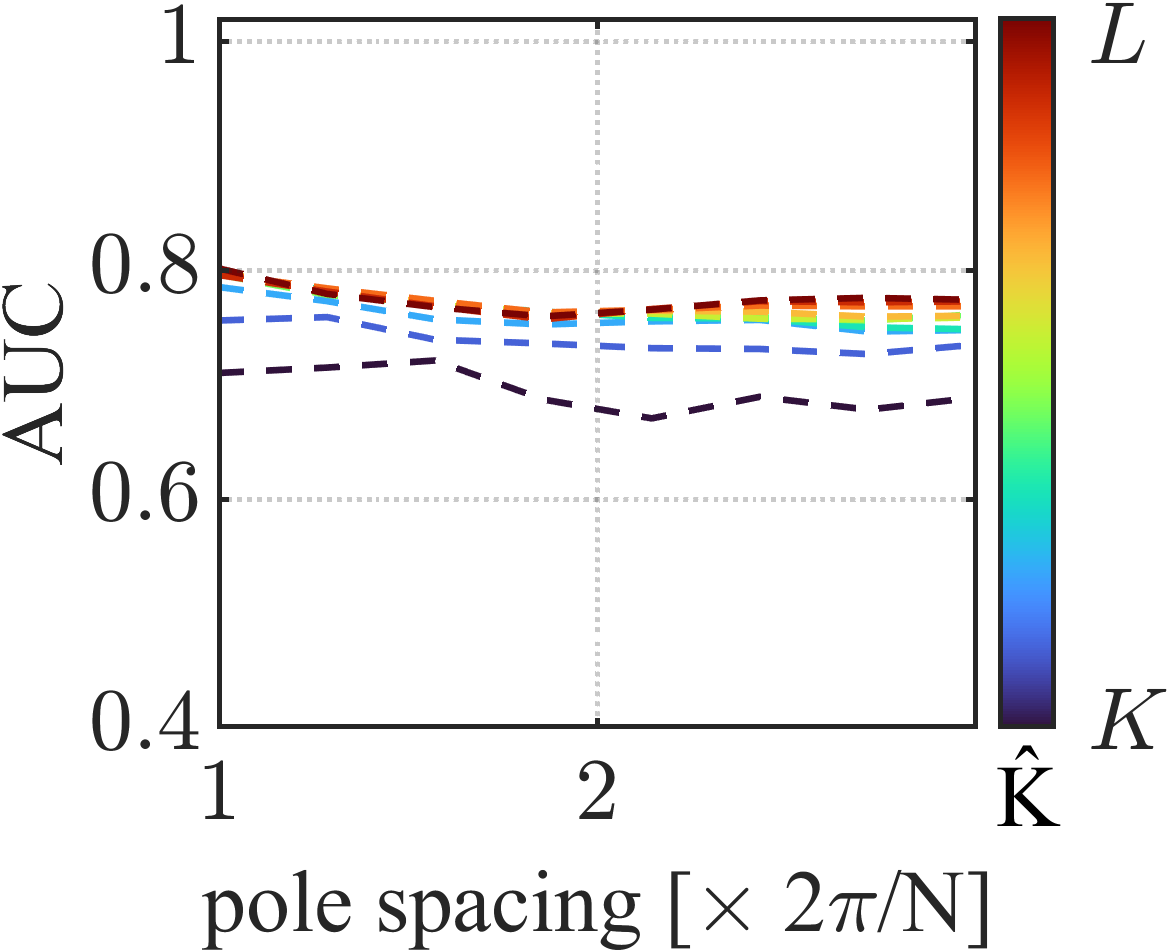}
        \caption{}
        \label{fig:sens_Khat_vs_minSepR}
    \end{subfigure}

    \par\smallskip

    \begin{subfigure}[t]{0.49\linewidth}
        \centering
        \includegraphics[width=\linewidth]{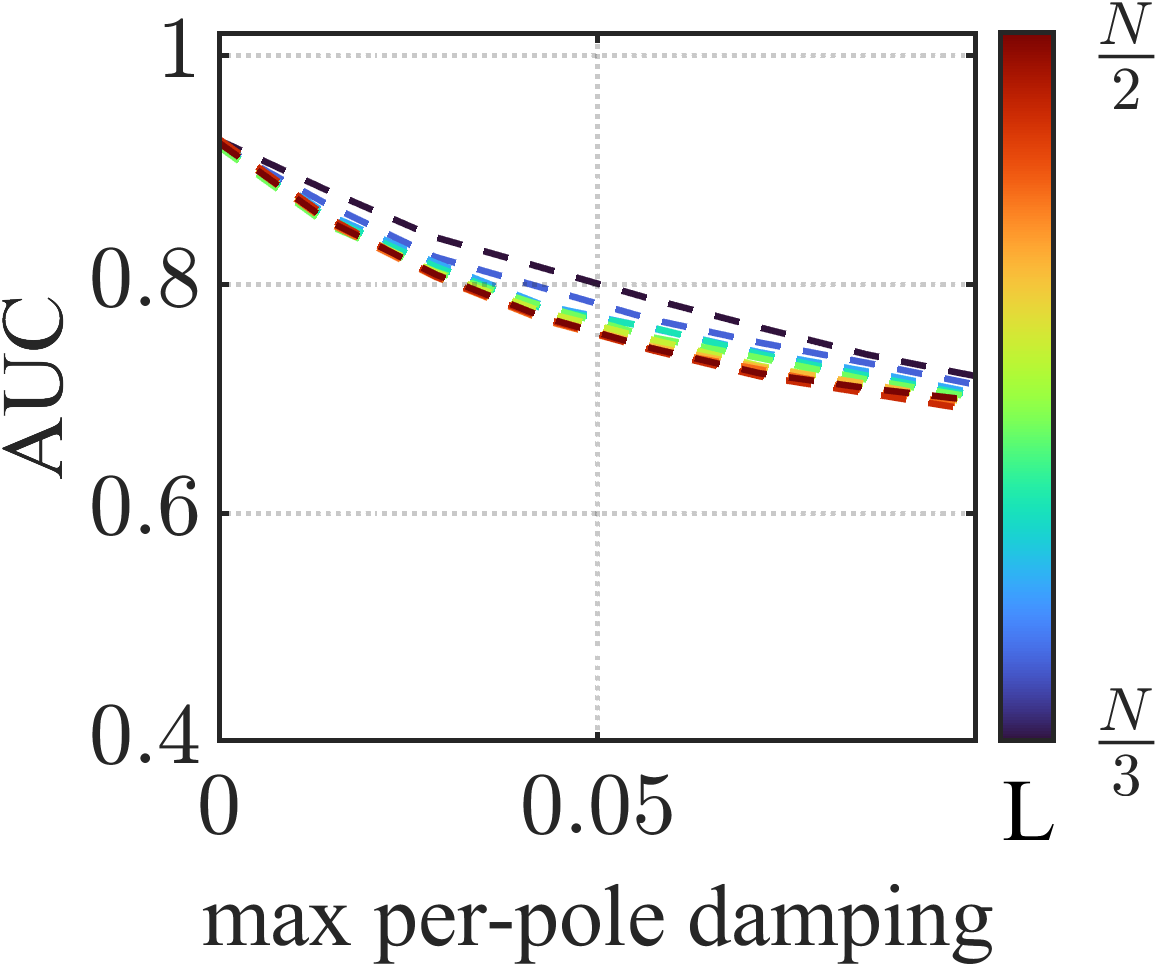}
        \caption{}
        \label{fig:sens_L_vs_damping}
    \end{subfigure}\hfill
    \begin{subfigure}[t]{0.49\linewidth}
        \centering
        \includegraphics[width=\linewidth]{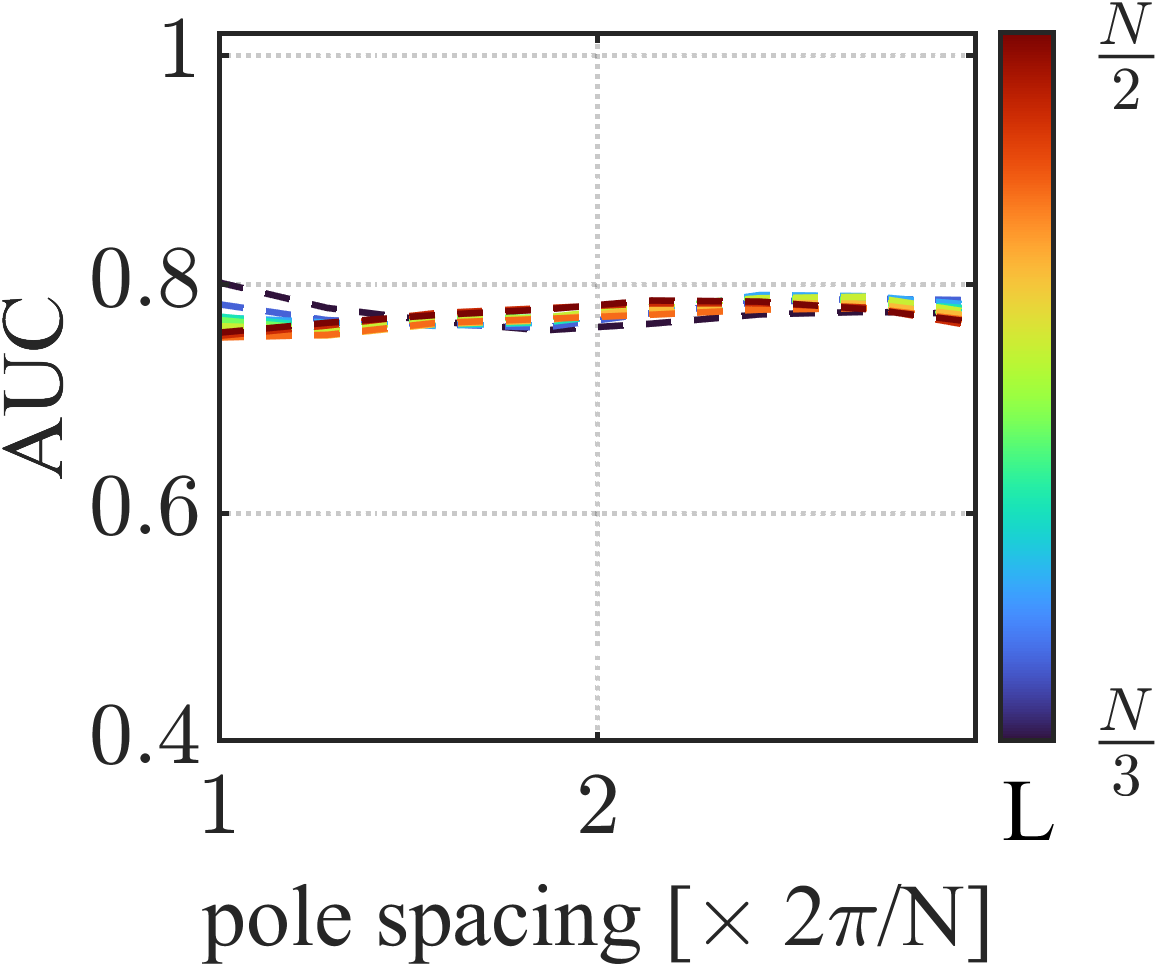}
        \caption{}
        \label{fig:sens_L_vs_minSepR}
    \end{subfigure}

    \caption{AUC-ROC versus maximum allowed pole damping  (left) and pole-spacing
    (right), varying $p$ (top), $\widehat K$ (middle), and $L$
    (bottom). Solid black curves correspond to selecting $p$ using the elbow method.}
    \label{fig:sensitivity_analysis_complementary}
\end{figure}

\begin{figure}[t]
    \centering

    \begin{subfigure}[t]{0.49\linewidth}
        \centering
        \includegraphics[width=1.05\linewidth]{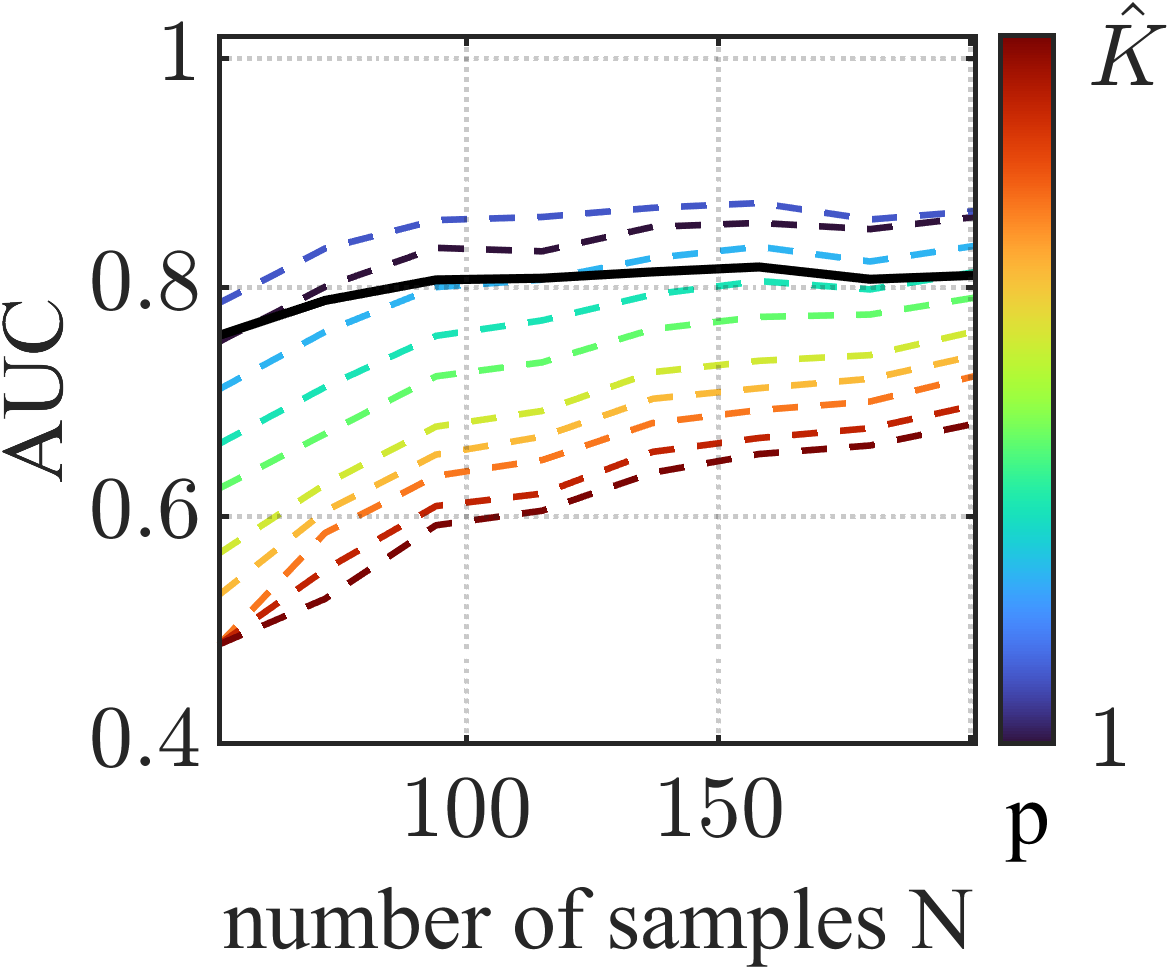}
        \caption{}
        \label{fig:sens_p_vs_N}
    \end{subfigure}\hfill
    \begin{subfigure}[t]{0.49\linewidth}
        \centering
        \includegraphics[width=1.05\linewidth]{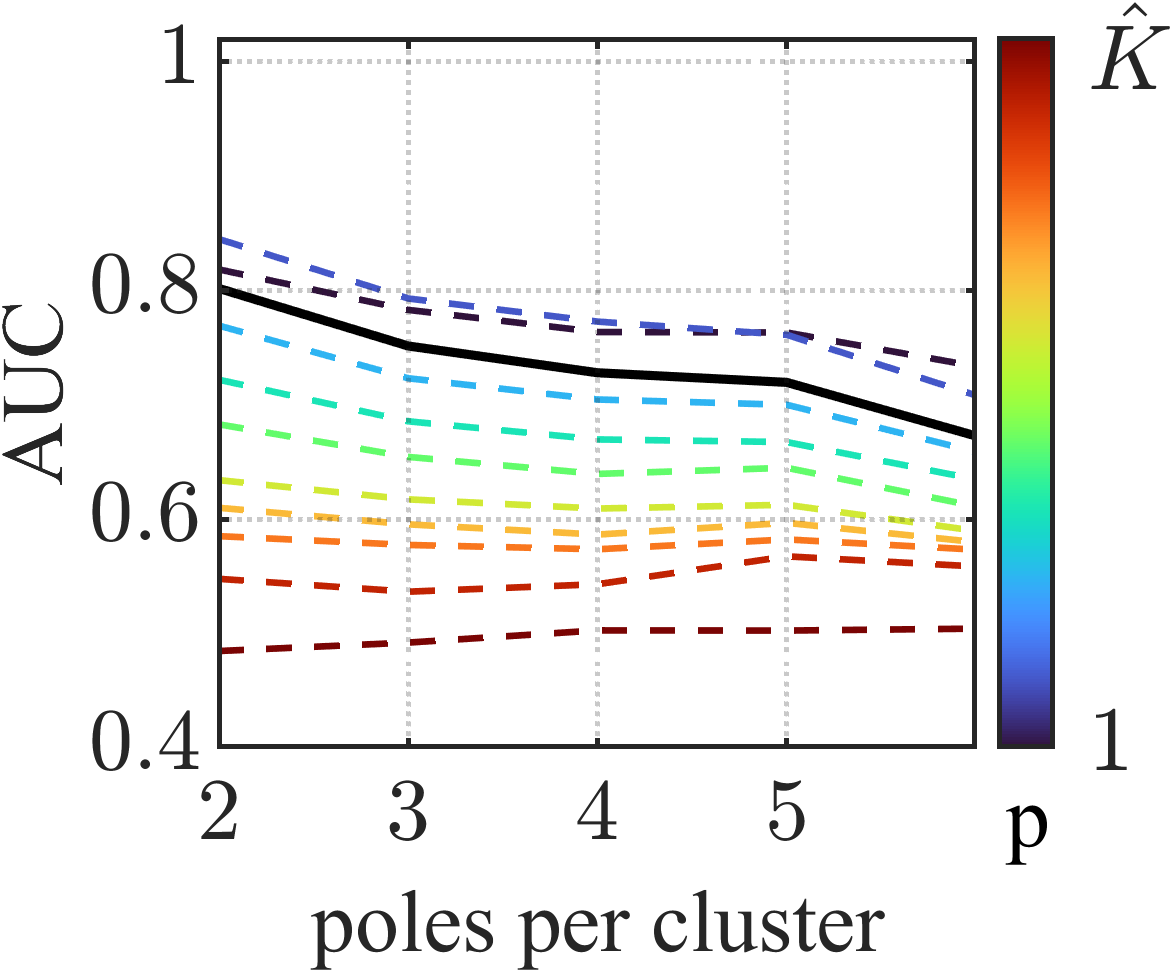}
        \caption{}
        \label{fig:sens_p_vs_polesPerCluster}
    \end{subfigure}

    \par\smallskip

    \begin{subfigure}[t]{0.49\linewidth}
        \centering
        \includegraphics[width=\linewidth]{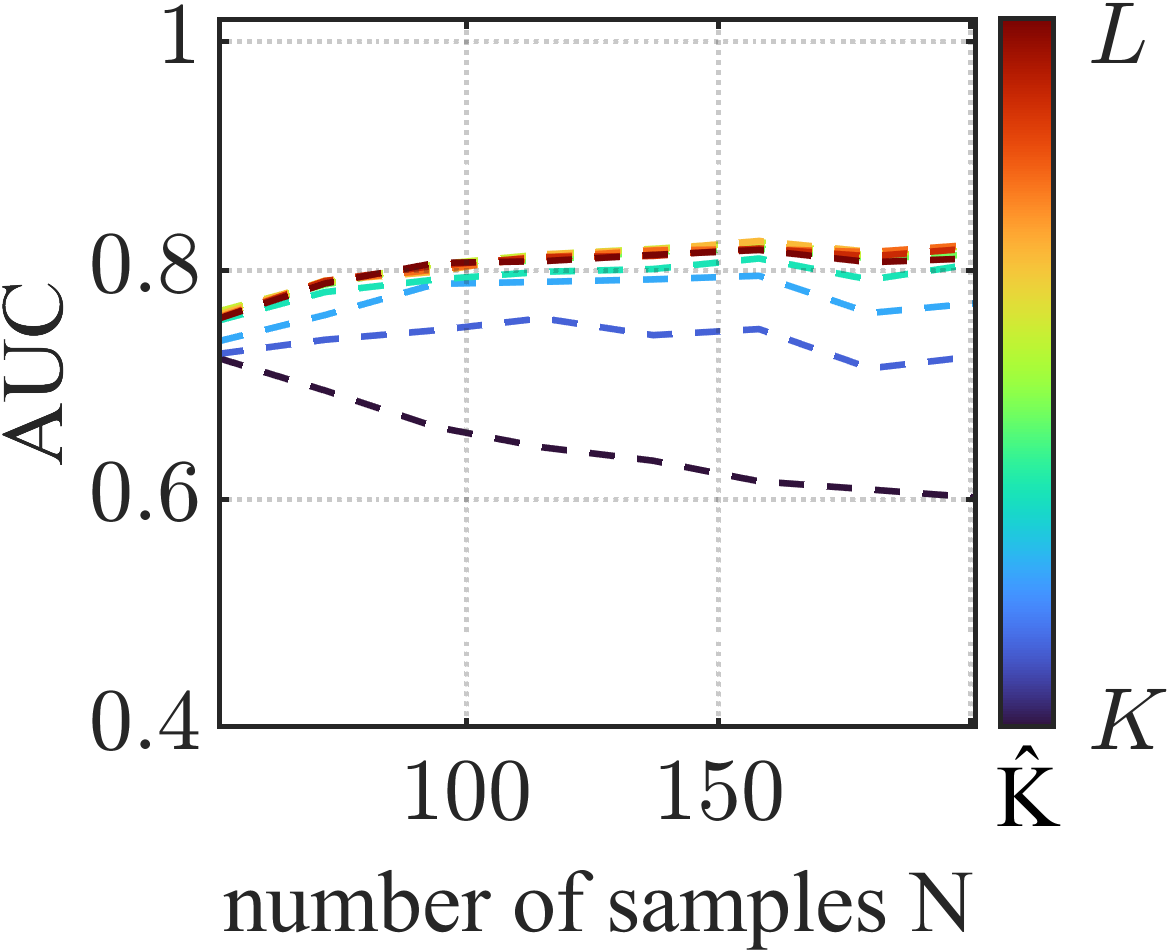}
        \caption{}
        \label{fig:sens_Khat_vs_N}
    \end{subfigure}\hfill
    \begin{subfigure}[t]{0.49\linewidth}
        \centering
        \includegraphics[width=\linewidth]{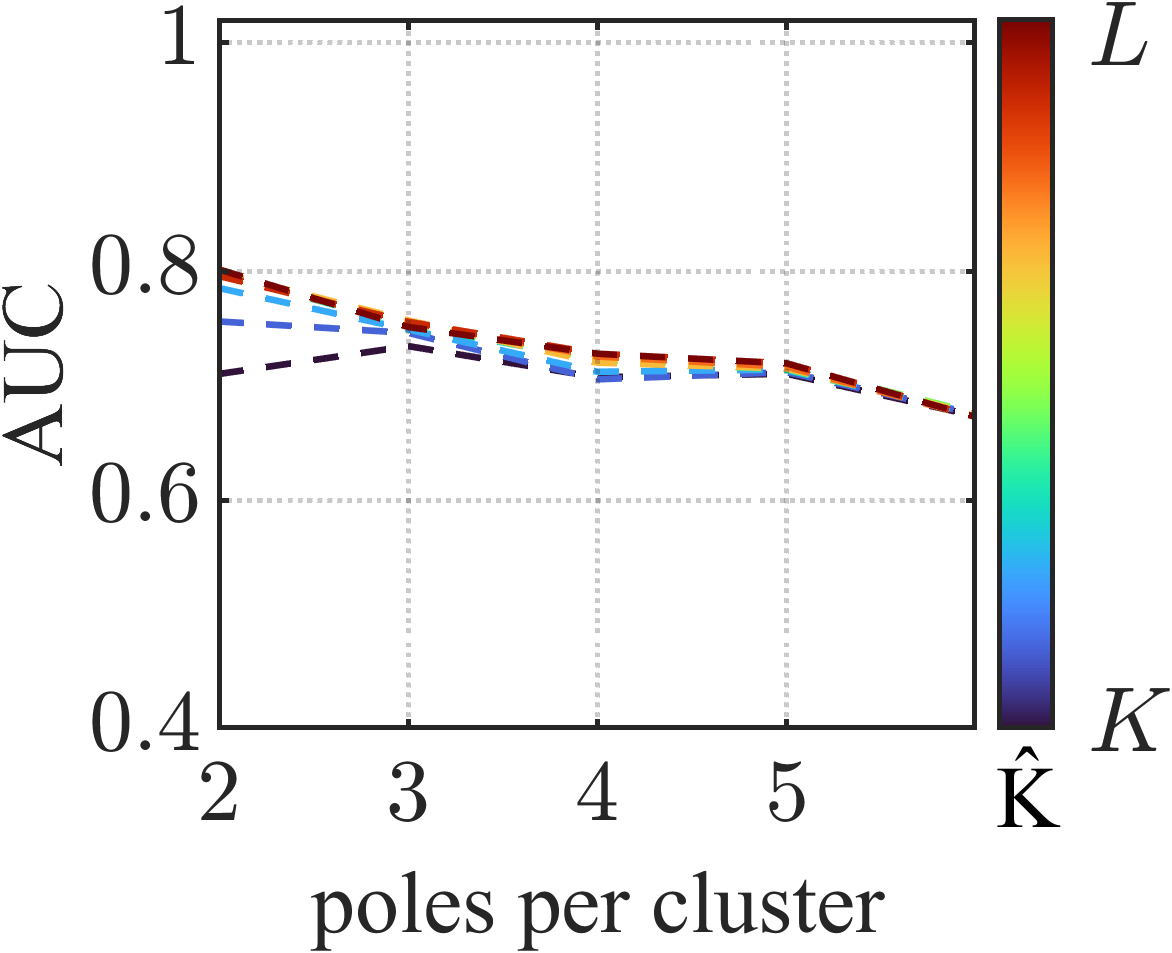}
        \caption{}
        \label{fig:sens_Khat_vs_polerPerCluster}
    \end{subfigure}

    \par\smallskip

    \begin{subfigure}[t]{0.49\linewidth}
        \centering
        \includegraphics[width=\linewidth]{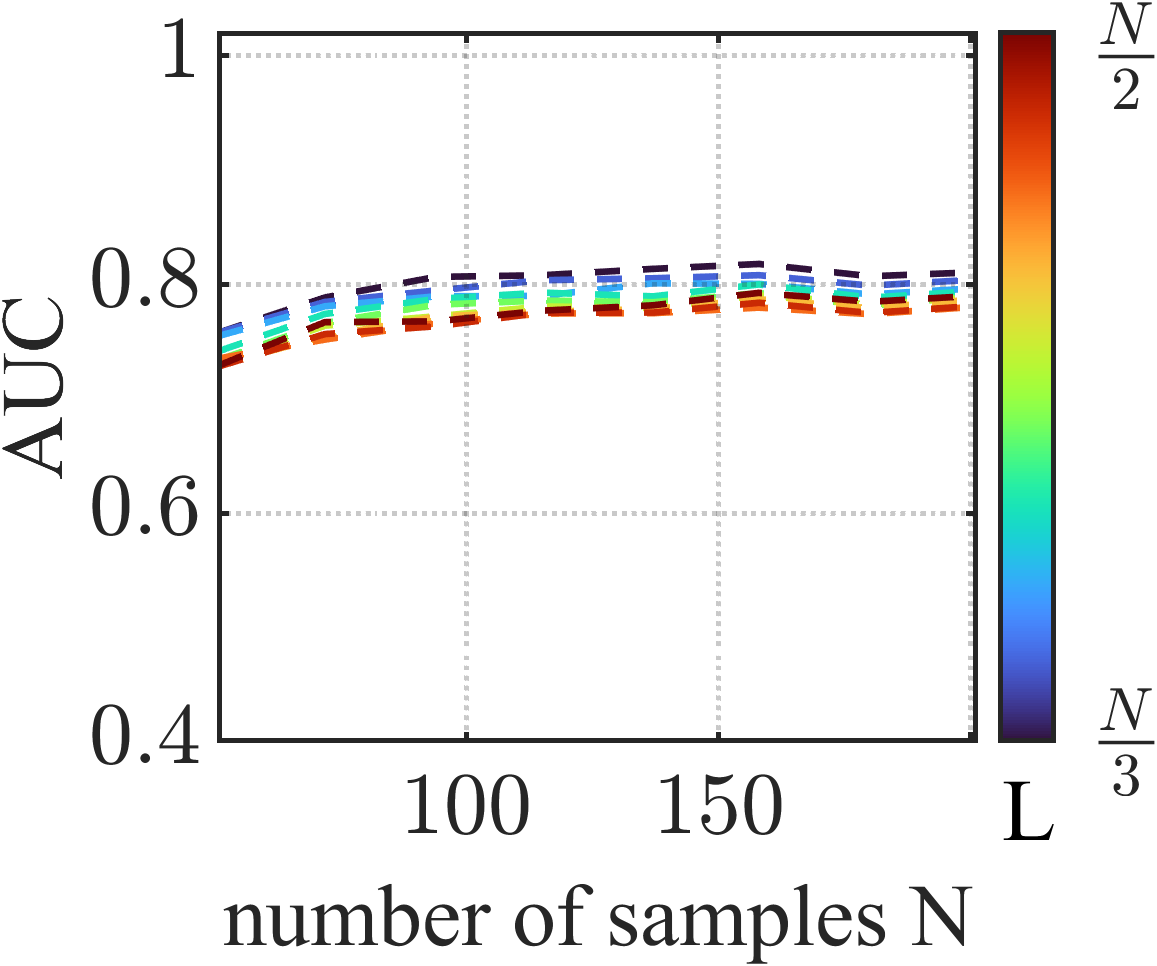}
        \caption{}
        \label{fig:sens_L_vs_N}
    \end{subfigure}\hfill
    \begin{subfigure}[t]{0.49\linewidth}
        \centering
        \includegraphics[width=\linewidth]{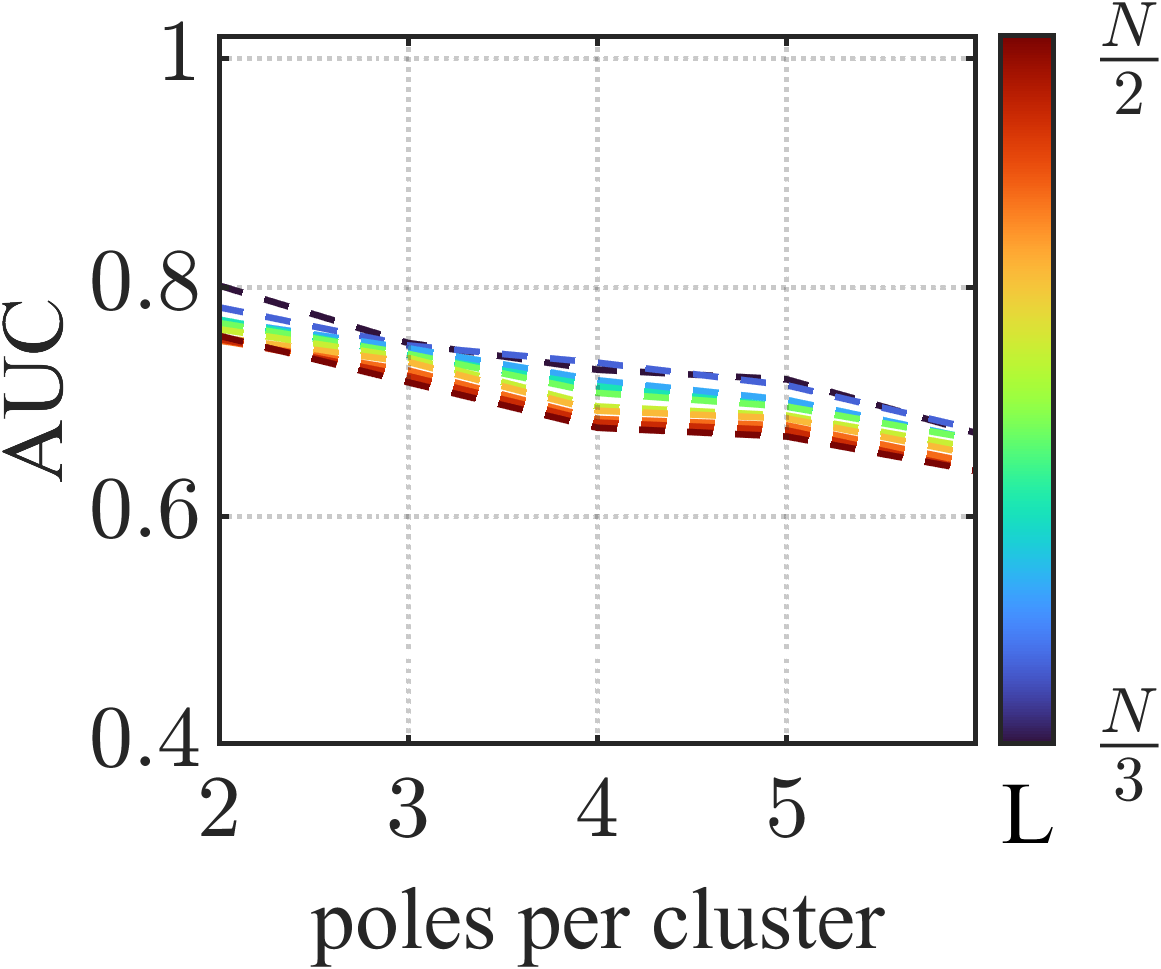}
        \caption{}
        \label{fig:sens_L_vs_polesPerCluster}
    \end{subfigure}

    \caption{AUC-ROC versus number of samples (left) and number of poles per cluster
    (right), varying $p$ (top), $\widehat K$ (middle), and $L$
    (bottom). Solid black curves correspond to selecting $p$ using the elbow method.}
    \label{fig:sensitivity_analysis_complementary2}
\end{figure}

\printbibliography
\end{refsection}
\end{NoHyper}
}

\end{document}